\documentclass[pra, showpacs, twocolumn, a4paper, nofootinbib]{revtex4}
\usepackage{graphicx}
\usepackage{amsmath, amsfonts, amssymb, bm}
\usepackage{amsmath}
\usepackage{verbatim}
\usepackage{bm}

\begin{document}

\newcommand{\ket}[1]{| #1 \rangle}
\newcommand{\bra}[1]{\langle #1 |}

\title{Spectral properties of a thresholdless dressed-atom laser}

\author{Min \surname{Luo}$^{a,b}$}
\author{Gao-xiang \surname{Li}$^{a}$}
\email{gaox@phy.ccnu.edu.cn}
\author{Zbigniew \surname{Ficek}$^{c}$}
\affiliation{$^{a}$Department of Physics, Huazhong Normal University, Wuhan 430079, China\\
$^{b}$Department of Physics, Hubei University of Nationalities, 445000, Enshi, Hubei, China\\
$^{c}$The National Centre for Mathematics and Physics, KACST, Riyadh 11442, P.O. Box 6086, Saudi Arabia}

\begin{abstract}
We investigate spectral properties of the atomic fluorescence and the output field of the cavity-mode of a single-atom dressed-state laser in a photonic crystal. We pay a particular attention to the behavior of the spectra in the presence of the frequency dependent reservoir and search for signatures of the thresholdless lasing. Although the thresholdless behavior has been predicted by analyzing the photon statistics of the cavity field, we find that the threshold behavior still exists in the spectrum of the cavity field. We find that the structure of cavity field spectrum depends strongly on the strange of the pumping rate. For low pumping rates, the spectrum is not monochromatic, it is composed of a set of discrete lines reveling the discrete (quantum) structure of the combined dressed-atom plus the cavity field system. We find that for a certain value of the pumping rate, the multi-peak structure converts into a single very narrow line centered at the cavity field frequency. A physical explanation of the behavior of the spectra is provided in terms of dressed states of the system. 
\end{abstract}

\maketitle

\section{Introduction}

The spectrum of the radiation field emitted by a system of atoms interacting with an electromagnetic field is known to provide fundamental insight into the energy structure and properties of the combined (entangled) atom-field system~\cite{sz97}. The spectral distribution of light emitted by coherently driven atoms simultaneously coupled to a multi-mode field has been the subject of numerous investigations over the years. The subject received its initial stimulus in the famous paper by Mollow~\cite{m69}, in which he showed that the fluorescence spectrum of a strongly driven two-level atom has a three-peaked structure, quite different from the spectrum expected for weak-field resonance fluorescence~\cite{km76}. The radiation properties of atoms interacting with a single rather than multi-mode electromagnetic field were also investigated in the context of a search for signatures of quantum effects in the matter-radiation interaction~\cite{be94}. The single-mode interaction, also known as the Jaynes-Cummings model, can be realized by confining the atoms within an optical cavity. The atom-cavity field coupling leads to "dressed" states, the eigenstates of the Hamiltonian of the atoms plus the cavity field, that serve as the basis energy states of the system~\cite{ct92}. The physical origin of the spectral features observed in the radiation field emitted by the system is clearly explained the dressed-atom model which provides the positions, intensities and widths of the spectral features in analytic form. In the dressed-atom approach, the radiation field is quantized which, in addition to the convenient studies of the spectral features, offers a potential setting for investigations of statistical and quantum features of the system. Many novel quantum effects have been predicted in observed, such as atomic collapses and revivals~\cite{sm83}, vacuum Rabi splitting~\cite{ag84,cb89}, photon antibunching and squeezing~\cite{rt91,th95}. The spectral properties of strongly driven atoms inside a single-mode cavity were also investigated in context of searching for signatures of two-photon lasing~\cite{gw92}, the photon-number distribution of the cavity field~\cite{qf93} and sub-natural linewidths~\cite{lg97,fq93}. 

The possibility of modifying linewidths of the spectral lines of the fluorescence field emitted by a three-level $V$-type atom using intense laser fields was first noted by Lorenzo Narducci and his colleagues~\cite{nso90}. They have found that the linewidth of the spectral lines on a given atomic transition can be switched into the linewidth of the other, much narrow atomic transition.
The phenomenon, termed as a dynamical narrowing of the spectral lines, can also be best understood in terms of the dressed-states of the system. The laser fields driving the atomic transitions lead to a coherent mixing of the atomic energy levels resulting in dressed states of the system. The transition rate between any two dressed states is proportional to the absolute square of the dipole transition moment connecting these two states. Since the dressed states depend on the Rabi frequencies of the driving fields, this results in the transition rates dependent on the Rabi frequencies and the spontaneous emission rates of the atomic transitions. Thus, the transition rates can be changed by changing the ratio between the Rabi frequencies.
Basically, the coherent mixing of the atomic energy levels modifies quantum fluctuations on the driven atomic transitions. Thus, the observed subnatural linewidth of the spectral line is the signature of the fluctuations stabilization in a driven atom by a coherent mixing of the atomic energy levels. In this context, the phenomenon of the dynamical narrowing of the spectral lines, predicted by Narducci et al. offers further interesting prospects for study of practical schemes for sub-natural spectral resolution and stabilization of the quantum fluctuations~\cite{gzm91}.

Narrowing of the spectral lines can also be achieved by placing the driven system inside a frequency dependent reservoir, the so called, tailored vacuum~\cite{kk95,lm88}. Transition rates between dressed states of the driven system can be modified with a suitable choosing of the Rabi frequency of the driving field. If the Rabi frequency is chosen to be much larger than the bandwidth of the reservoir field, some of the dressed-atom frequencies can then be found outside the bandwidth of the reservoir field. As a result, transitions at these frequencies are not affected by spontaneous emission. The selective decoupling of the dressed-atom transition frequencies may lead to narrowing and even suppression of the spectral lines of the fluorescence.

There have also been a considerable interest in the generation of lasing action between dressed states of strongly driven two-level atoms~\cite{zl91,lz90,lb91}. The concept of the dressed-atom laser has been introduced, where the lasing action can occur even in the absence of population inversion between the bare states of the atoms. This kind of lasing is often referred to as lasing without population inversion. In fact, the lasing occurs with population inversion which exists between the dressed states involved in the lasing transition.

Recently, the spectral and statistical properties of the single-atom dressed-state laser in a photonic band-gap material were shown to be qualitatively different from that arising from the ordinary frequency independent reservoir~\cite{fj04,fl06,tl08}. The band-gap material serves as a frequency dependent reservoir, where the density of the electromagnetic field modes varies sharply with frequency. Furthermore, it has been shown that in the presence of the photonic material, the dressed-atom system can operate as a thresholdless laser~\cite{ll09}. 

In this paper, we examine the atomic fluorescence and the cavity field output spectra for features indicative of the thresholdless lasing which results in the single-atom dressed-state laser from suppression of spontaneous emission on the lasing transition~\cite{ll09}. Our interest in this particular problem stems from the fact that the thresholdless behavior of the system has been predicted by analyzing the photon statistics of the cavity field~\cite{rc94,tp94,dm88,bk94,pd99,sh99}. It is interesting to consider if a thresholdless behavior could also be observed in the spectra.

The predicted suppression of the threshold behavior in the statistics of the cavity photons occurs if a photonic band-gap material is employed to reduce the density of the electromagnetic field modes in the lasing region of the transition. As is well known, a strongly driven two-level atom fluorescence at three distinct frequencies, the driving laser frequency and at two sideband frequencies shifted symmetrically from the laser frequency by the Rabi frequency. 
We consider an off-resonance driving field and the cavity mode tuned to resonance with the lower frequency Rabi sideband of the dressed atom. Our treatment is based on the master equation of the reduced density operator of the driven atom plus the cavity field, which we derive by combining two previously used approaches, the Keitel-Knight-Narducci-Scully~\cite{kk95} approach to the dynamics of a driven atom in a frequency dependent reservoir, and Freedhoff-Quang~\cite{fq93} approach to the dynamics of a driven atom coupled simultaneously to a single-mode cavity field and a frequency independent reservoir. In the absence of the band-gap material, we distinguish three regimes of the pumping rate, below, near and above threshold, at which the spectra exhibit significantly different structures. Below the threshold, the spectrum exhibits two peaks characteristic of the vacuum Rabi splitting. Near the threshold, a multi-peak structure is observed indicating a successive excitation of the discrete energy states of the system. Above the threshold, the cavity field spectrum is composed of a single very narrow peak, whereas the atomic fluorescence spectrum exhibits a Mollow-like triplet. 
In the presence of the band-gap material, the spectra display interesting modifications. First of all, we can distinguish only two ranges of the pumping rate at which the spectra have different character. A multi-peak structure emerges at relatively low pumping rates. The spectra exhibit a multi-peak structures even in the limit of vanishing pumping. Secondly, we find that the spectral lines are significantly narrower than that predicted in the absence of the band-gap material. We observe an interesting feature that in the case of the cavity field spectrum, the multi-peak structure builds inside the vacuum Rabi doublet, whereas the fluorescence spectrum the structure builds inside as well as outside the doublet. Finally, we demonstarte that the cavity field spectrum exhibits a threshold behavior that at a certain pumping rate, the multi-peak spectrum converts into a single very narrow peak located at the cavity mode frequency. The threshold behavior exists despite the fact that there is no threshold behavior in the photon statistics of the cavity field.

The presence of the threshold raises a question of the meaning of a thresholdless laser, whether the thresholdless operation of the system should be determined by either, the photon statistics or the spectrum of the cavity field, or both.

\section{The dressed-atom system}

We consider a system consisting of an atom located within a single-mode cavity engineered inside a photonic crystal. The atom is modeled as a two-level system with lower level $\ket 1$ and upper level $\ket 2$ separated by energy $\hbar\omega_{0}$, and driven by an intense single-mode laser field of the resonant Rabi frequency $2\Omega_{0}$. The laser field propagates transverse to the cavity axis and is treated classically, while the cavity mode is treated as a quantum field in a state $\ket n$ containing $n$ photons of frequency $\omega_{c}$.  

The reason for engineering the cavity inside a photonic crystal is twofold. On the one hand, the photonic crystal provides a system for a strong coupling of the atom to a previlage mode. On the other hand, the crystal provides a reservoir of a multi-mode electromagnetic field whose modal density varies sharply on a frequency scale much smaller than the Rabi frequency of the driving field~\cite{wj04,ya87,jo87,jq94}. Theoretically, the variation of the density of the field modes with frequency is expressed in terms of a step function that varies sharply at the cut-off frequency $\omega_{b}$. 

The dynamics of the total system; the driven atom plus the cavity plus the reservoir, is determined by the density operator $\rho_{T}$, which in the interaction picture satisfies the master equation 
\begin{eqnarray}
\frac{\partial }{\partial t}\tilde{\rho}_{T} = -\frac{i}{\hbar}\left[\tilde{H},\tilde{\rho}_{T} \right] +\frac{1}{2}\kappa{\cal L}_{c}\tilde{\rho}_{T} ,\label{e1}
\end{eqnarray}
where
\begin{eqnarray}
\tilde{\rho}_{T} = \exp\left(-\frac{i}{\hbar}H^{\prime}t\right)\rho_{T}\exp\left(\frac{i}{\hbar}H^{\prime}t\right) \label{e2}
\end{eqnarray}
is the density operator of the system in the interaction picture, with
\begin{eqnarray}
H^{\prime} = \hbar\omega_{L} a^\dagger a +\frac{1}{2}\hbar\omega_{L}\!\left(\sigma_{22} -\sigma_{11}\right) + \hbar\sum_{k}\omega_{L}a_{k}^{\dag}a_{k} ,\label{e3}
\end{eqnarray}
and
\begin{eqnarray}
{\cal L}_{c}\tilde{\rho}_{T} = \left(2a\tilde{\rho}_{T}{a}^\dag-{a}^\dag{a}\tilde{\rho}_{T} -\tilde{\rho}_{T}{a}^\dag{a}\right) \label{e4}
\end{eqnarray}
is an operator representing the damping of the cavity field with the rate $\kappa$.

The Hamiltonian $\tilde{H}$, appearing in Eq.~(\ref{e1}), is composed of three terms
\begin{eqnarray}
\tilde{H} = \tilde{H}_{0}+\tilde{H}_{L}+\tilde{H}_{I} ,\label{e5}
\end{eqnarray}
where
\begin{eqnarray}
\tilde{H}_{0} = \frac{1}{2}\hbar\Delta_{a}\left(\sigma_{22} -\sigma_{11}\right) + \hbar\Delta_{c}a^{\dag}a +\hbar\sum_{k}\Delta_{k}a_{k}^{\dag}a_{k} \label{e6}
\end{eqnarray}
is the free Hamiltonian of the atom, the cavity field, and the multi-mode reservoir, respectively,
\begin{eqnarray}
\tilde{H}_{L}= \hbar\Omega_{0}\left(\sigma_{12}+\sigma_{21}\right) \label{e7}
\end{eqnarray}
is the interaction Hamiltonian of the driving laser field with the atom, and
\begin{eqnarray}
\tilde{H}_{I} = i\hbar g\left(a^{\dag}\sigma_{12}-\sigma_{21}a\right)
+i\hbar\sum_{k}g_{k}\left(a^{\dag}_{k}\sigma_{12}-\sigma_{21}a_{k}\right) \label{e8}
\end{eqnarray}
is the interaction Hamiltonian of the privilege cavity mode and the remaining reservoir modes with the atom. 

In Eqs.~(\ref{e1})-(\ref{e8}), the parameters $g$ and $g_{k}$ describe the coupling constants between the atom and the cavity mode and the reservoir modes, respectively, $a\ (a^{\dag})$ and $a_{k}\ (a^{\dag}_{k})$ are, respectively, the cavity-mode and the radiation reservoir annihilation (creation) operators, $\sigma_{ij} =\ket i \bra j \ (i,j=1,2)$ are the bare atomic operators, and $2\Omega_{0}$ is the resonant Rabi frequency of the laser field which, for simplicity, we have assumed is real and positive. The remaining parameters $\Delta_{a}=\omega_{a}-\omega_{L}$, $\Delta_{c}=\omega_{c}-\omega_{L}$, and $\Delta_{k}=\omega_{k}-\omega_{L}$ are the detunings of the atomic resonance frequency $\omega_{0}$, the cavity-mode frequency $\omega_{c}$, and the frequency $\omega_{k}$ of the $k$th mode of the radiation reservoir from the frequency $\omega_{L}$ of the driving laser field. 

There are three fields interacting with the atom, the laser field whose the coupling strength to the atom is given by the Rabi frequency $\Omega_{0}$, the cavity field coupled to the atom with the strength $g$ and the multi-mode vacuum reservoir  whose the $k$th mode is coupled to the atom with the strength $g_{k}$. We shall assume that the coupling $\Omega_{0}$ dominates over $g$ that, on the other hand, is much stronger than $g_{k}$, so that $\Omega_{0}\gg g\gg g_{k}$. We shall follow the following procedure, first "dressing" the atom in the laser field and then couple the resulting dressed-atom system to the cavity and the reservoir fields.

As we shall be interested in the spectral properties of the radiation field produced by a dressed-state laser, we make the customary transformation of representing the master equation in terms of dressed states of the driven atom~\cite{ct92}. The dressed states are the eigenstates of the Hamiltonian $\tilde{H}_{AL}= \frac{1}{2}\hbar\Delta_{a}\left(\sigma_{22} -\sigma_{11}\right)+\tilde{H}_{L}$. Since the driving field is treated classically in our calculations, the diagonalization of the Hamiltonian $\tilde{H}_{AL}$ produces the so-called semiclassical dressed states
\begin{eqnarray}
\ket{\tilde{1}} = \cos\phi\ket 1 -\sin\phi\ket 2 ,\quad \ket{\tilde{2}} = \sin\phi\ket 1 +\cos\phi\ket 2 ,\label{e9}
\end{eqnarray}
where $ \cos^{2}\phi = \frac{1}{2}\left[1+\Delta_{a}/(2\Omega)\right]$ and $2\Omega=(4\Omega_{0}^{2}+\Delta_{a}^{2})^{1/2}$ is the Rabi frequency of the detuned field. 

We now introduce dressed-state operators, $R_{ij}=\ket{\tilde{i}}\bra{\tilde{j}}$, and find that the Hamiltonian $\tilde{H}_{0}+\tilde{H}_{L}$ becomes
\begin{eqnarray}
\tilde{H}_{0} +\tilde{H}_{L} \equiv \tilde{H}_{d} = \hbar\Delta_{c}a^{\dag}a +\hbar\Omega R_{3}+\sum_{\lambda}\Delta_{\lambda}a_{\lambda}^{\dag}a_{\lambda} ,\label{e10}
\end{eqnarray}
where $R_{3}=R_{22}-R_{11}$. Note that the Hamiltonian (\ref{e10}) is diagonal in the basis of the product states $\ket{\tilde{i}}\otimes\ket n \otimes\ket{n_{k}}$, which prompts us to make the unitary transformation
\begin{eqnarray}
\bar{\rho}_{T} = \exp\left(-\frac{i}{\hbar}H_{d}\, t\right)\tilde{\rho}_{T}\exp\left(\frac{i}{\hbar}H_{d}\, t\right) ,\label{e11}
\end{eqnarray}
which results in the density operator in the dressed-atom picture. It follows from Eqs.~(\ref{e1}) and~(\ref{e11}) that the maser equation for the density operator in the dressed-atom picture becomes
\begin{eqnarray}
\frac{\partial }{\partial t}\bar{\rho}_{T}\left( t\right) = -\frac{i}{\hbar}\left[\bar{H}_{I},\bar{\rho}_{T}\left( t\right) \right] ,\label{e12}
\end{eqnarray}
where 
\begin{align}
\bar{H}_{I} &= g\, a^{\dag}\!\left[\frac{1}{2}\sin(2\phi)R_{3}\,{\rm e}^{i\Delta_{c}t}+\cos^{2}\!\phi R_{12}\,{\rm e}^{i(\Delta_{c}-2\Omega)t}\right.\nonumber\\
&\left. -\sin^{2}\!\phi R_{21}\,{\rm e}^{i(\Delta_{c}+2\Omega)t}\right]
+\sum_{k}g_{k}\, a_{k}^{\dag}\!\left[\frac{1}{2}\sin(2\phi)R_{3}\,{\rm e}^{i\Delta_{k}t}\right.\nonumber\\
&\left. +\cos^{2}\!\phi R_{12}\,{\rm e}^{i(\Delta_{k}-2\Omega)t}-\sin^{2}\!\phi R_{21}\,{\rm e}^{i(\Delta_{k}+2\Omega)t}\right] + {\rm H.c.} \label{e13}
\end{align}
Note that in the dressed-atom picture the evolution of the density operator is determined only by the interaction Hamiltonian (\ref{e8}).

We now proceed to eliminate the reservoir field by tracing the density operator of the total system over the space of the reservoir modes to obtain the reduced density operator of the dressed atom and the cavity mode only. Since the multi-mode field appears as a frequency dependent reservoir, we assume that the Rabi frequency is much larger than the width of the band edge of the band-gap material is much smaller than the spontaneous emission rate of the atomic bare transition and the dressed-atom transition frequencies $\omega_{L}$ and $\omega_{L}\pm 2\Omega$ are far away from the the band-edge frequency $\omega_{b}$. In this case, it is possible to obtain a master equation for the reduced density operator consistent with the Born-Markov approximation. In addition, we assume that the cavity mode is tuned to exact resonance with the low-frequency Rabi sideband, i.e. $\Delta_{c}=-2\Omega$. With this frequency tuning, we can distinguish that certain terms in the master equation are stationary in time and the other oscillating at frequencies $2\Omega$ and $4\Omega$. Since the Rabi frequency is large, these oscillatory terms make a negligible contribution to the dynamics of the system and we may ignore them. This simplification is a form of the rotating-wave approximation. With such approximations, and after using the standard procedure of eliminating of the reservoir modes, we find that the reduced density operator $\rho$ of the dressed atom plus the cavity mode has the form
\begin{align}
\frac{\partial{\rho}}{\partial{t}} &= g_{1}\!\left[\left(R_{12}a-{a}^\dag{R_{21}}\right),\rho\right] +\frac{1}{2}\kappa{\cal L}_{c}\rho \nonumber\\
&+\frac{1}{8}\gamma_{0}\!\left(2R_3\rho{R_3}-\rho{R_3^2}-R_3^2\rho\right) \nonumber\\
&+\frac{1}{2}\gamma_{-}\!\left(2R_{21}\rho{R_{12}}-R_{12}R_{21}\rho-\rho{R_{12}R_{21}}\right) \nonumber\\
&+\frac{1}{2}\gamma_{+}\!\left(2R_{12}\rho{R_{21}}-R_{21}R_{12}\rho-\rho{R_{21}R_{12}}\right) ,\label{e14}
\end{align}
where $g_1 = g\sin^{2}\phi$ is the "effective" coupling constant of the dressed atom to the cavity field,
\begin{eqnarray}
\gamma_{0} &=& \gamma \sin^{2}(2\phi) u(\omega_{L}-\omega_{b}) ,\ \
\gamma_{-} = \gamma \sin^{4}(\phi) u(\omega_{-}-\omega_{b})  ,\nonumber \\
\gamma_{+} &=& \gamma \cos^{4}(\phi) u(\omega_{+} -\omega_{b}) ,\label{e15}
\end{eqnarray}
are the damping rates between the dressed states of the system, and $u(\omega_{i} -\omega_{b}),\, \omega_{i}=\omega_{L},\omega_{\pm}$, is the unit step function describing the transfer function of the band-gap material. The function is evaluated at three frequencies corresponding to the three transition frequencies between the dressed states, $\omega_{i} =\omega_{L}$ and $\omega_{i} =\omega_{\pm}=\omega_{L}\pm 2\Omega$. Since the frequencies $\omega <\omega_{b}$ are forbidden in the band-gap material, this opens a possibility to eliminate spontaneous transitions at selected frequencies of the dressed-atom system.  

The master equation (\ref{e14}) is very similar in form to that one found for the the frequency independent reservoir. In particular, the damping rates $\gamma_{0}, \gamma_{+}$ and $\gamma_{-}$ differ only in that the damping rate $\gamma$ of the bare atomic transition is replaced by $\gamma u(\omega_{i} -\omega_{b})$, the frequency dependent rates. Thus, modifications to the master equation brought by the band-gap material are essentially reflected in the appearance of three, frequency dependent, damping rates. As we shall see, the modifications, although look trivial at the first glance, will in fact significantly affect the dynamics of the system and the spectral distribution of the emitted light.

The damping rate $\gamma_{+}$ corresponds to spontaneous emission from the upper dressed
state $\ket{\tilde{2}}$ to the lower dressed state $\ket{\tilde{1}}$ of the manifold below and occurs at frequency $\omega_{+}=\omega_{L}+2\Omega$, whereas the damping rate $\gamma_{-}$ corresponds to spontaneous emission from the lower dressed state to the upper dressed state of the manifold below and occurs at frequency $\omega_{-}=\omega_{L}-2\Omega$.
It should be noted that the rate $\gamma_{-}$ appears as the decay rate on the transition resonant to the cavity frequency,  whereas the rate~$\gamma_{+}$, that appears as a damping rate plays, in fact, the role of an incoherent pumping of the dressed system from $|\tilde{2}\rangle$ to $|\tilde{1}\rangle$. In other words, it is a pure incoherent pumping process that transfers the population to the upper state of the lasing transition.
Thus, by a suitable tuning of the dressed-atom frequency $\omega_{-}$ to the edge frequency $\omega_{b}$, one can eliminate spontaneous emission at the frequency of the cavity field.

We now introduce density-matrix elements with respect to the dressed states, denoting $\rho_{ij}= \bra{\tilde{i}}\rho\ket{\tilde{j}}$, and using the master equation (\ref{e14}), we obtain the following simple equations of motion for the populations and coherence
\begin{align}
\frac{\partial}{\partial{t}}{\rho_{11}} &= g_{1} \left(a\rho_{21}+\rho_{12}a^{\dagger}\right) -\gamma_{-}\rho_{11} +\gamma_{+}\rho_{22} +\frac{1}{2}\kappa{\cal L}_{c}\rho_{11} ,\nonumber\\
\frac{\partial}{\partial{t}}{\rho_{22}} &= -g_{1}\left(a^{\dagger}\rho_{12}+\rho_{21}a\right) -\gamma_{+}\rho_{22} +\gamma_{-}\rho_{11} +\frac{1}{2}\kappa{\cal L}_{c}\rho_{22} ,\nonumber\\
\frac{\partial}{\partial{t}}{\rho_{12}} &= g_{1}\left(a\rho_{22}-\rho_{11}a\right)
-\Gamma_{c}\,\rho_{12} +\frac{1}{2}\kappa{\cal L}_{c}\rho_{12} ,\nonumber\\
\frac{\partial}{\partial{t}}{\rho_{21}} &= g_{1}\left(\rho_{22}a^{\dagger}-a^{\dagger}\rho_{11}\right) 
-\Gamma_{c}\,\rho_{21} +\frac{1}{2}\kappa{\cal L}_{c}\rho_{21} ,\label{e16}
\end{align}
where $\Gamma_{c} = (\gamma_{0}+\gamma_{+} +\gamma_{-})/2$ is the decay rate associated with the $\rho_{12}$ coherence. 

Equations (\ref{e16}) are the basic equations for calculating the spectra of the atomic fluorescence and the cavity field. In the following we always assume that the Rabi frequency of the laser field is much larger than the coupling constant of the cavity mode and the damping rates of the atom and the cavity mode, $2\Omega \gg g,\gamma,\kappa$. Under such conditions, we can selectively tune the dressed-atom frequencies to the transmissive or to the forbidden band of the band-gap material.

\section{Atomic fluorescence and cavity field spectra}

We now follow the procedure of Freedhoff and Quang~\cite{fq93} to calculate the atomic fluorescence and cavity field spectra. The evaluation of the spectra simplifies by introducing 
the following Hermitian combination of the density matrix elements
\begin{eqnarray}
\rho^{(1)}&=& \rho_{22}+\rho_{11} ,\ \rho^{(2)} = \rho_{22}-\rho_{11} ,\nonumber\\
\rho^{(3)}&=& \frac{1}{2}\left(\rho_{21}a+a^{\dagger}\rho_{12}\right) ,\nonumber\\
\rho^{(4)}&=& \frac{1}{2}\left(a\rho_{21}+\rho_{12}a^{\dagger}\right) .\label{e17}
\end{eqnarray}
The advantage of working with the Hermitian combinations is the most evident if one considers the photon-number representation, ${\rho^{(i)}_{n,n+m}}\equiv\langle{n}|\rho^{(i)}|{n+m}\rangle$, where it can be easily found that the equations of motion for the diagonal matrix elements decouple from the equations of motion for the off-diagonal elements. 

Using Eq.~(\ref{e16}), we find that the equations of motion for the density matrix elements ${\rho^{(i)}_{n,n+m}}$ satisfy a set of coupled differential equations, which can be written in a matrix form as
\begin{eqnarray}
\frac{\partial}{\partial{t}}Z^{(m)}_{n} = {\bf A}^{(m)}_{n}Z^{(m)}_{n-1}+{\bf B}^{(m)}_{n}Z^{(m)}_{n}+{\bf C}^{(m)}_{n}Z^{(m)}_{n+1} ,\label{e18}
\end{eqnarray}
where $Z^{(m)}_{n}$ is the column vector composed of the density matrix elements, and ${\bf A}^{(m)}_{n}$, ${\bf B}^{(m)}_{n}$ and  ${\bf C}^{(m)}_{n}$ are $4\times 4$ matrices composed of the coefficients of the equation. The explicit forms of the vector $Z^{(m)}_{n}$ and the matrices ${\bf A}^{(m)}_{n}$, ${\bf B}^{(m)}_{n}$ and  ${\bf C}^{(m)}_{n}$ are given in  Appendix A.

Equation (\ref{e18}) is in the form of a vector recurrence relation which is solved by a continued-fraction method, and the solution can be written in terms of Green's function matrices $G^{(m)}_{n,j}(t)$~as
\begin{align}
Z^{(m)}_{n}(t) = \sum_{j=0}^{\infty}Z^{(m)}_{n,j}(t) = \sum_{j=0}^{\infty}Z^{(m)}_{j}(0)G^{(m)}_{n,j}(t) ,\label{e19}
\end{align}
where $Z^{(m)}_{j}(0)$ are the initial values of the density matrix elements.

The general solution (\ref{e19}) can be applied to evaluate the spectrum of light emitted by the system. Two kinds of radiation are produced by the system, the atomic fluorescence emitted out of the sides of the cavity and the cavity-mode field, leaking the cavity through the cavity mirrors.

\subsection{Spectral distribution of the atomic fluorescence}

In the dressed-atom picture and under the condition of well separated dressed-atom transition frequencies, the spectrum of the atomic fluorescence can be evaluated separately at each of the dressed-atom frequency as
\begin{eqnarray}
S_{a}(\omega) = S^{(-)}(\omega)+S^{(0)}(\omega) + S^{(+)}(\omega) ,\label{e20}
\end{eqnarray}
where
\begin{align}
S^{(-)}(\omega) &= \gamma_{-}\,{\rm Re}\!\int_{0}^{\infty}\!d\tau {\rm e}^{-i(\omega-\omega_{-})\tau}\langle{R_{12}(t+\tau)R_{21}(t)\rangle}_{s} ,\nonumber \\
S^{(0)}(\omega) &= \frac{1}{4}\gamma_{0}\,{\rm Re}\!\int_{0}^{\infty}\!d\tau {\rm e}^{-i(\omega-\omega_{L})\tau}
\langle{R_{3}(t+\tau)R_{3}(t)\rangle}_{s} ,\nonumber\\
S^{(+)}(\omega) &= \gamma_{+}\,{\rm Re}\!\int_{0}^{\infty}\!d\tau{\rm e}^{-i(\omega-\omega_{+})\tau} \langle{R_{21}(t+\tau)R_{12}(t)\rangle}_{s} ,\label{e21}
\end{align}
and the subscript $s$ denotes the steady-state value of the correlation function. The required spectrum therefore follows directly from the dressed-state two-time correlation functions, which can be evaluated from the general solution (\ref{e19}).

Since we are interested in the spectral distribution of the field emitted at the cavity frequency, which is tuned to the frequency of the lower Rabi sideband, we will therefore focus our attention on the spectrum at the frequency $\omega_{-}=\omega_{L}-2\Omega$. The two-time correlation functions appearing in Eq.~(\ref{e21}) are found by applying the quantum regression theorem to the equation of motion (\ref{e16}). The general solution for the spectrum, given in terms of the Green function matrix is of the form
\begin{align}
&S^{(-)}(\omega) = \gamma_{-}\left\{\frac{\Gamma_{c}\langle{R_{12}R_{21}}\rangle_{s}}{\Gamma_{c}^{2} +\nu^{2}}\right. \nonumber\\
&\left. +{\rm Re}\!\sum^{\infty}_{n,j=0}\frac{g_1\sqrt{n+1}}{\Gamma_{c} +i\nu}\left[G^{(1)}_{n,j}(i\nu)Z^{(1)}_{j}(0)\right]_{2}\right\} ,
\end{align}
where $[G^{(1)}_{n,j}(i\nu)Z^{(1)}_{j}(0)]_{2} = \rho^{(2)}_{n,n+1}(i\nu)$ is the Laplace transformation of $\rho^{(2)}_{n,n+1}(\tau)$, and $\nu =\omega -\omega_{-}$.

\subsection{Spectral distribution of the cavity field}

The incoherent part of the spectrum of the cavity field evaluated at the frequency $\omega_{-}=\omega_{L}-2\Omega$ is defined as the Fourier transformation of the two-time correlation function of the fluctuation operators 
\begin{align}
S_{c}(\omega) = 2{\rm Re}\int^{\infty}_{0}\!d\tau\, {\rm e}^{i(\omega-\omega_{-})\tau}
\langle\delta a^{\dag}(t+\tau) \delta a(t) \rangle_{s} ,
\end{align}
where $\delta a(t) = a(t) -\langle a(t)\rangle$. 

Using the quantum regression theorem, one can show that the two-time correlation function of the field fluctuation operators can be expressed in terms of the density matrix elements as
\begin{eqnarray}
\langle\delta a^{\dag}(t+\tau) \delta a(t) \rangle_{s} = \sum^{\infty}_{n=0}\sqrt{n+1}\rho^{(1)}_{n,n+1}(\tau) ,\label{e24}
\end{eqnarray}
where $\rho^{(1)}_{n,n+1}(\tau)$ is the solution of Eq.~(\ref{e16}) with the initial condition given by the steady-state values of the density matrix elements. 

The spectrum is obtained by taking the Fourier transform of Eq.~(\ref{e24}), or equivalently it can be done replacing $\tau$ by $\tau = i\nu$. Hence,  we readily find the following formula for the cavity field spectrum
\begin{align}
S_{c}(\omega) = 2{\rm Re}\sum^{\infty}_{n,j=0}\sqrt{n+1}\left[G^{(1)}_{n,j}(-i\nu) Z^{(1)}_{j}(0)\right]_{1} .\label{e25}
\end{align}

In the following section, we present numerical results for the atomic fluorescence and the cavity field spectra evaluated for different values of the pumping rate of the lasing transition.

\section{Numerical results}

\begin{figure}[t]
\includegraphics[width=5cm,keepaspectratio,clip]{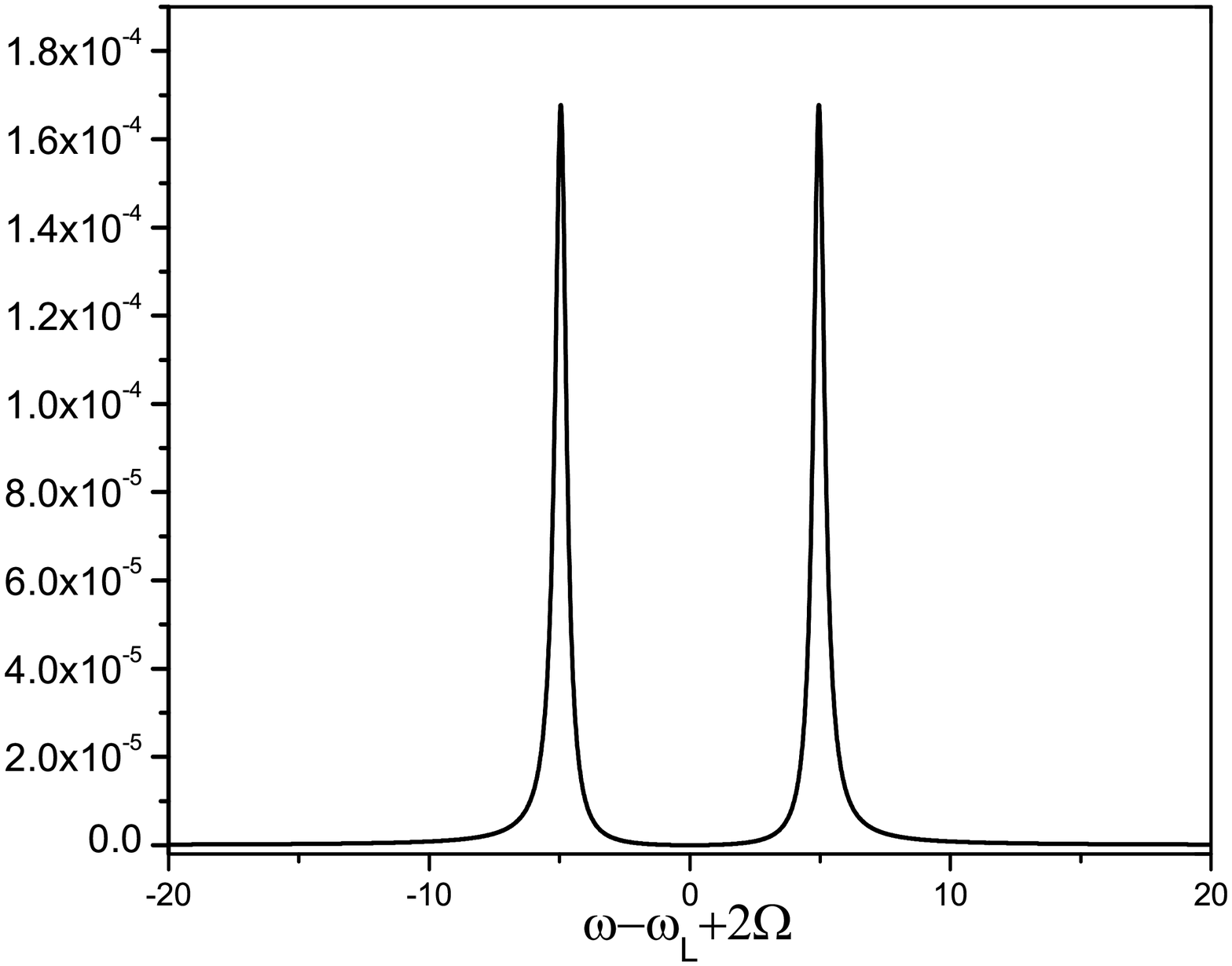}\\
\includegraphics[width=5cm,keepaspectratio,clip]{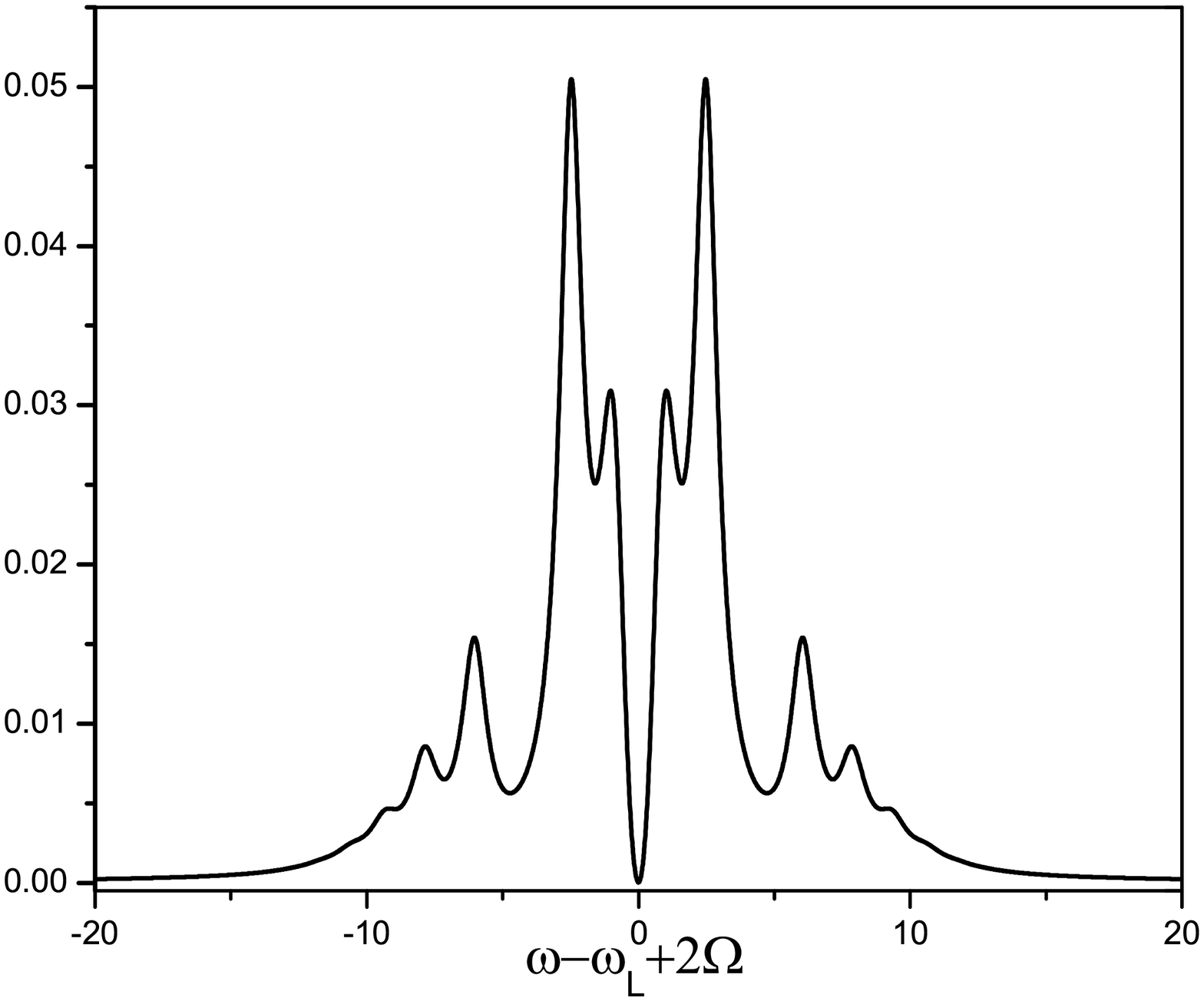}\\
\includegraphics[width=5cm,keepaspectratio,clip]{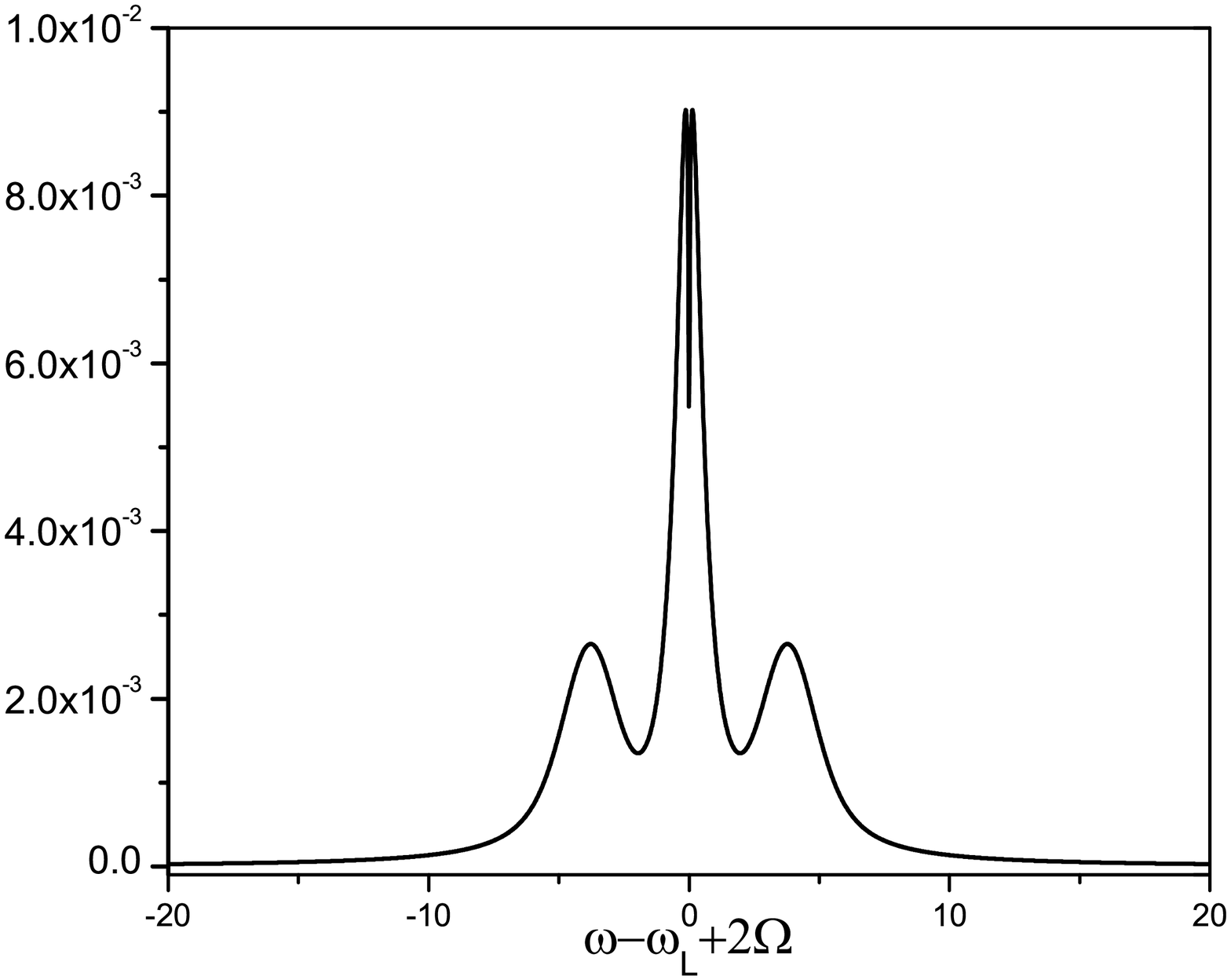}\\
\includegraphics[width=5cm,keepaspectratio,clip]{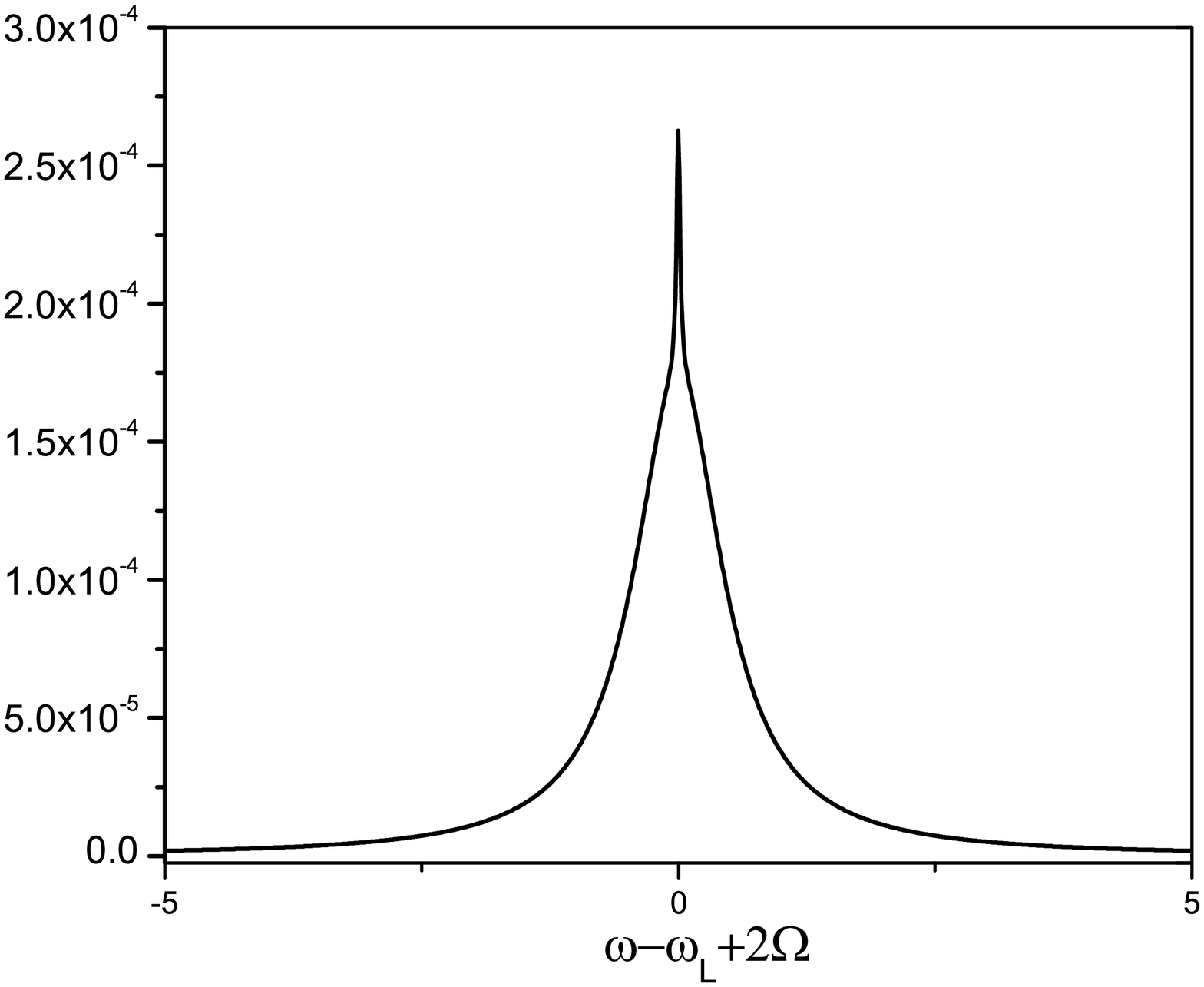}\\
\caption{The atomic fluorescence spectrum $S^{(-)}(\omega)$ as a function of frequency $\omega -\omega_{L}+2\Omega$ for $\kappa = 0.05$, $g=5$, $\gamma_{-}\neq 0$ and different $\delta_{a}$: (a) $\delta_{a}=-5$, (b) $\delta_{a}= 0$, (c) $\delta_{a}= 1$, (d) $\delta_{a}= 5$. All the parameters are normalized to $\gamma=1$.} \label{fig1}
\end{figure}

\begin{figure}[hbp]
\includegraphics[width=5cm,keepaspectratio,clip]{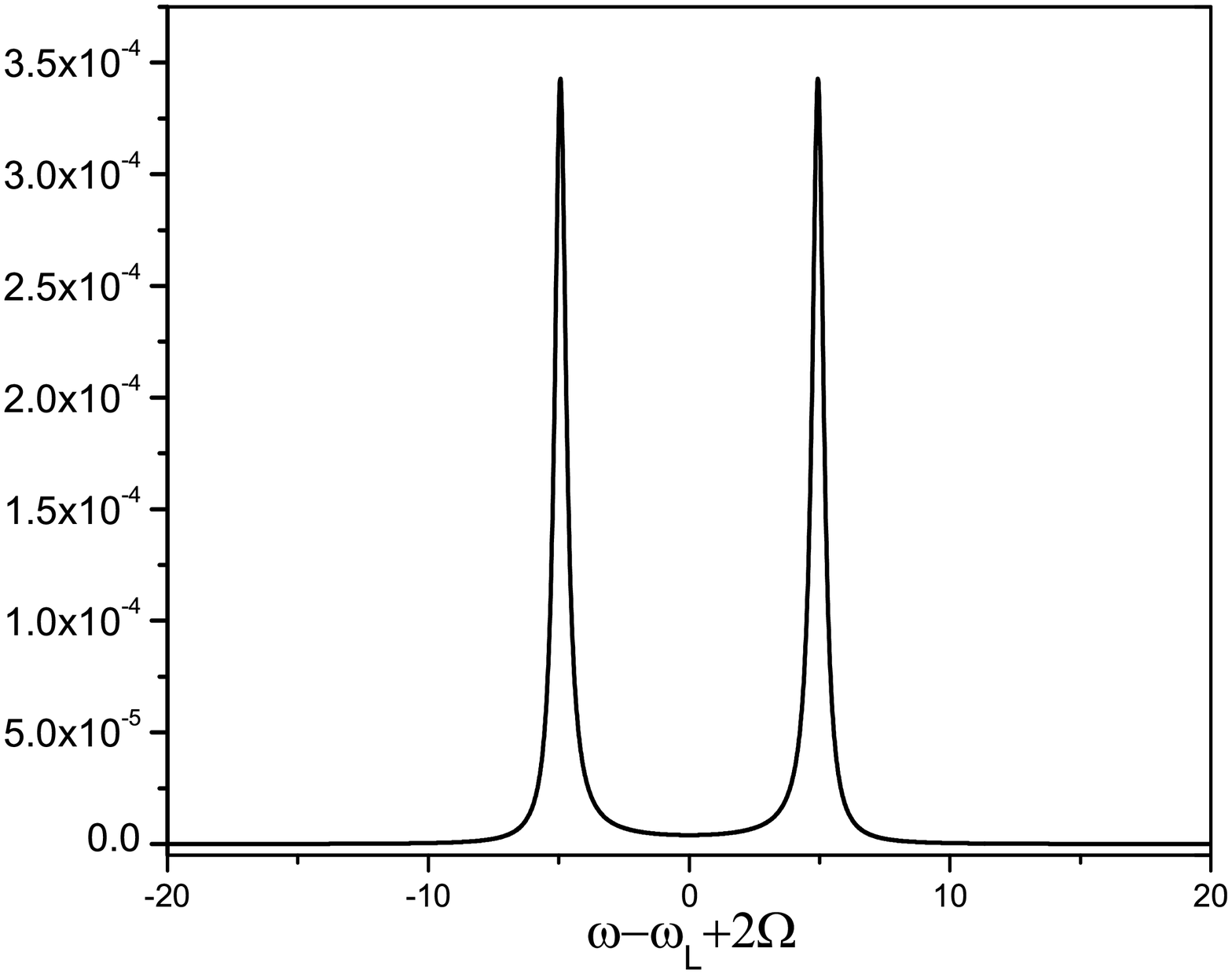}\\
\includegraphics[width=5cm,keepaspectratio,clip]{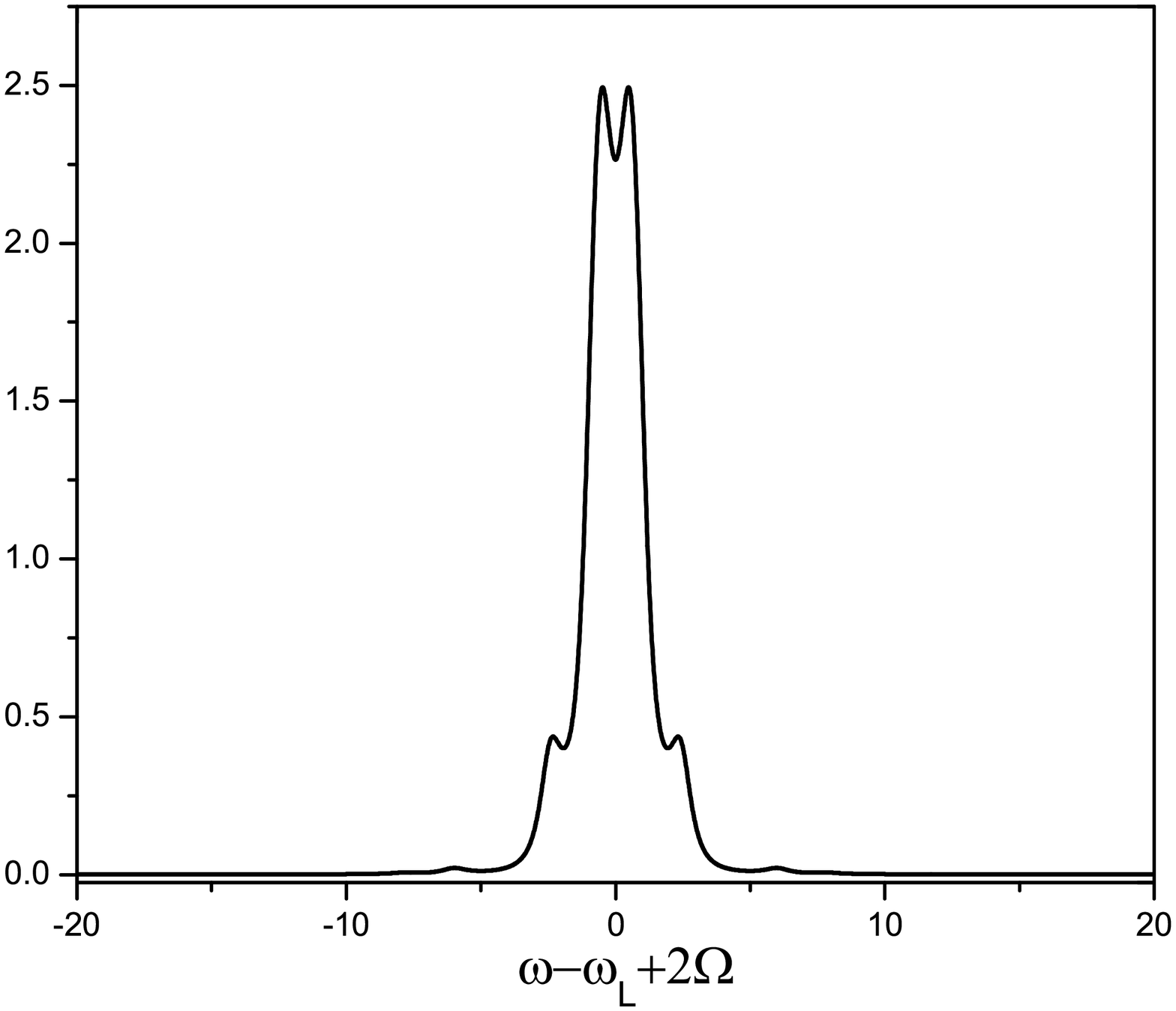}\\
\includegraphics[width=5cm,keepaspectratio,clip]{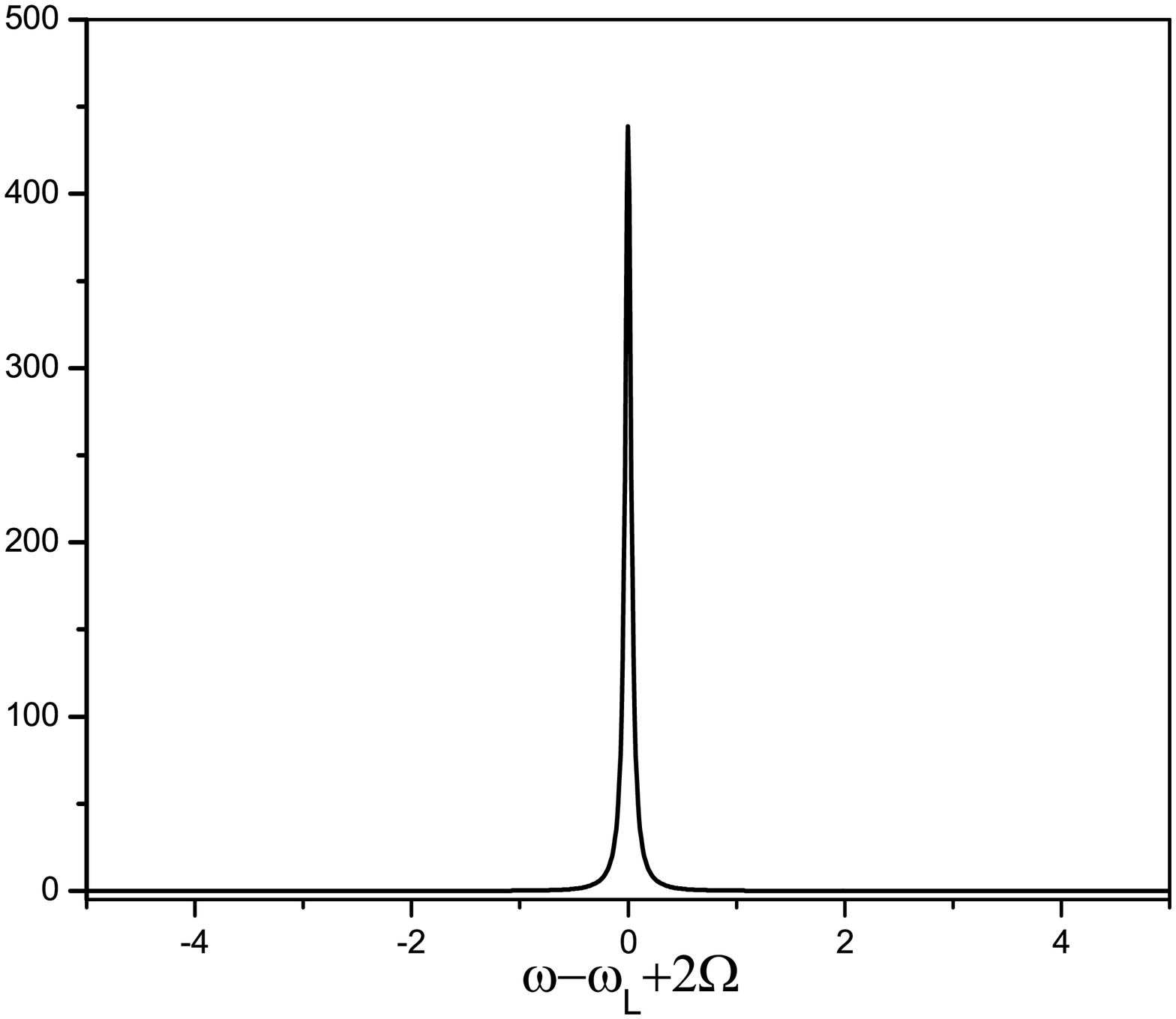}\\
\includegraphics[width=5cm,keepaspectratio,clip]{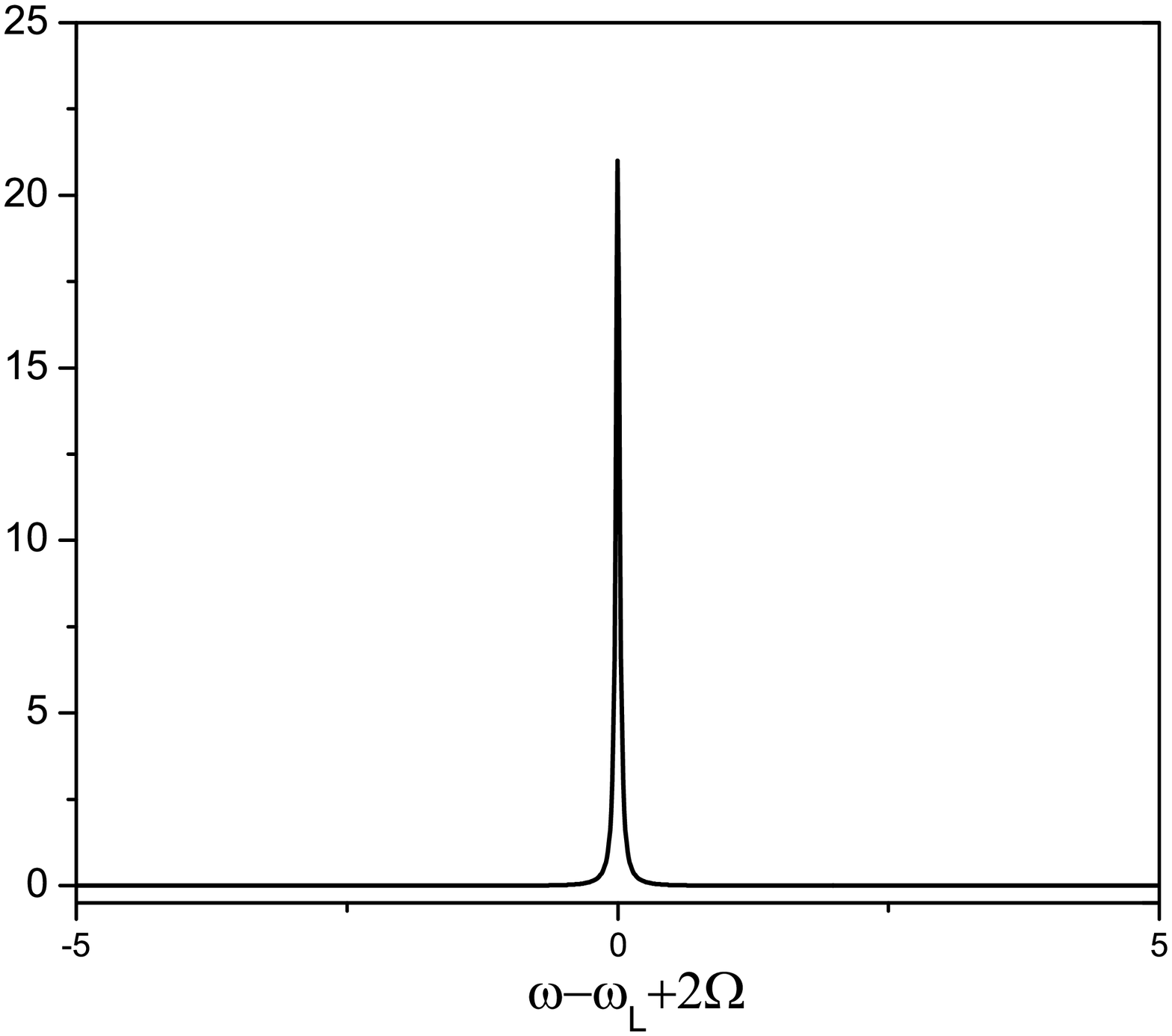}\\
\caption{The cavity field spectrum $S_{c}(\omega)$ as a function of frequency $\omega -\omega_{L}+2\Omega$ for $\kappa = 0.05$, $g=5$, $\gamma_{-}\neq 0$ and different $\delta_{a}$: (a) $\delta_{a}=-5$, (b) $\delta_{a}= 0$, (c) $\delta_{a}= 1$, (d) $\delta_{a}= 5$. All the parameters are normalized to $\gamma=1$.} \label{fig2}
\end{figure}

\begin{figure}[hbp]
\includegraphics[width=5cm,keepaspectratio,clip]{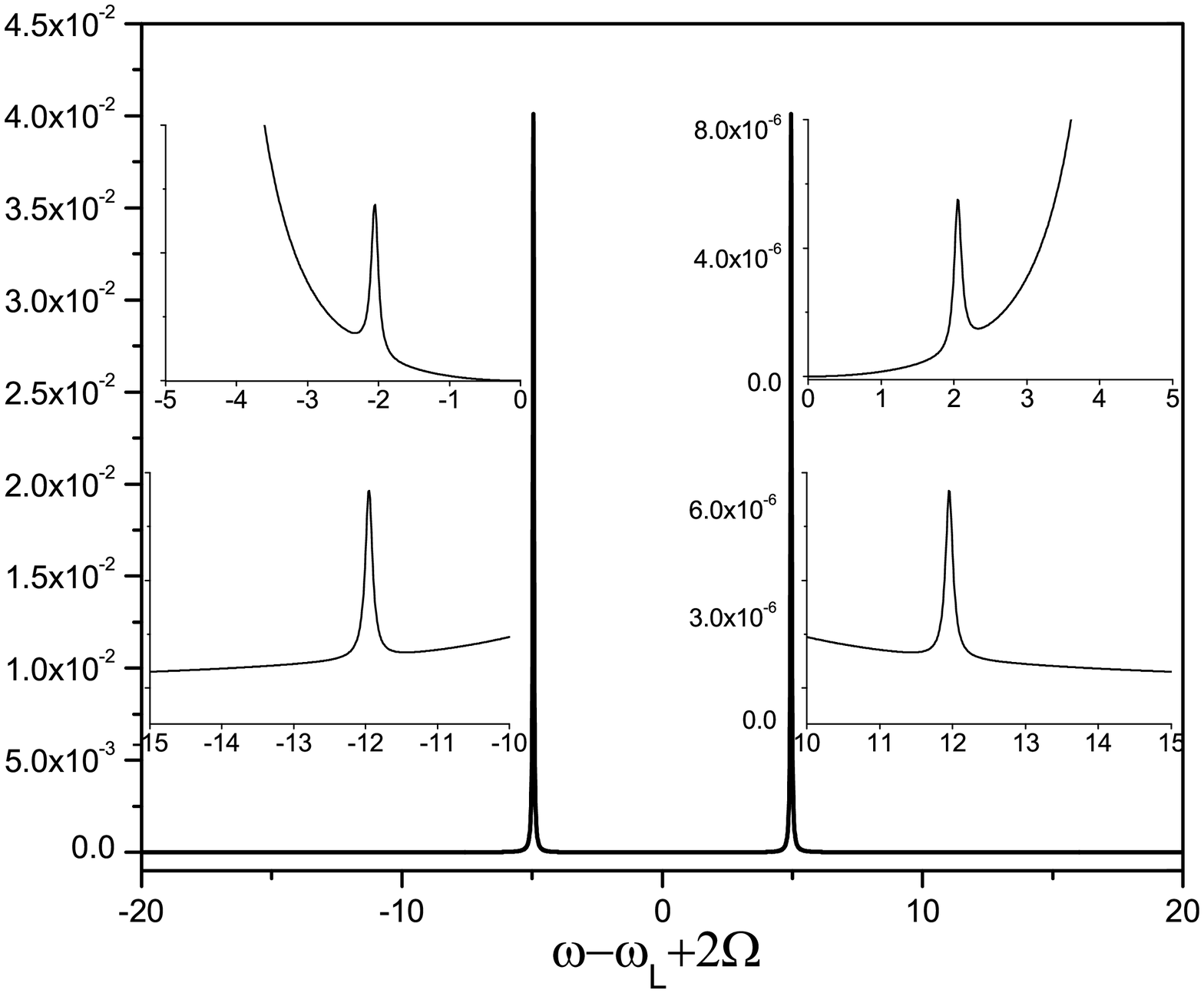}\\
\includegraphics[width=5cm,keepaspectratio,clip]{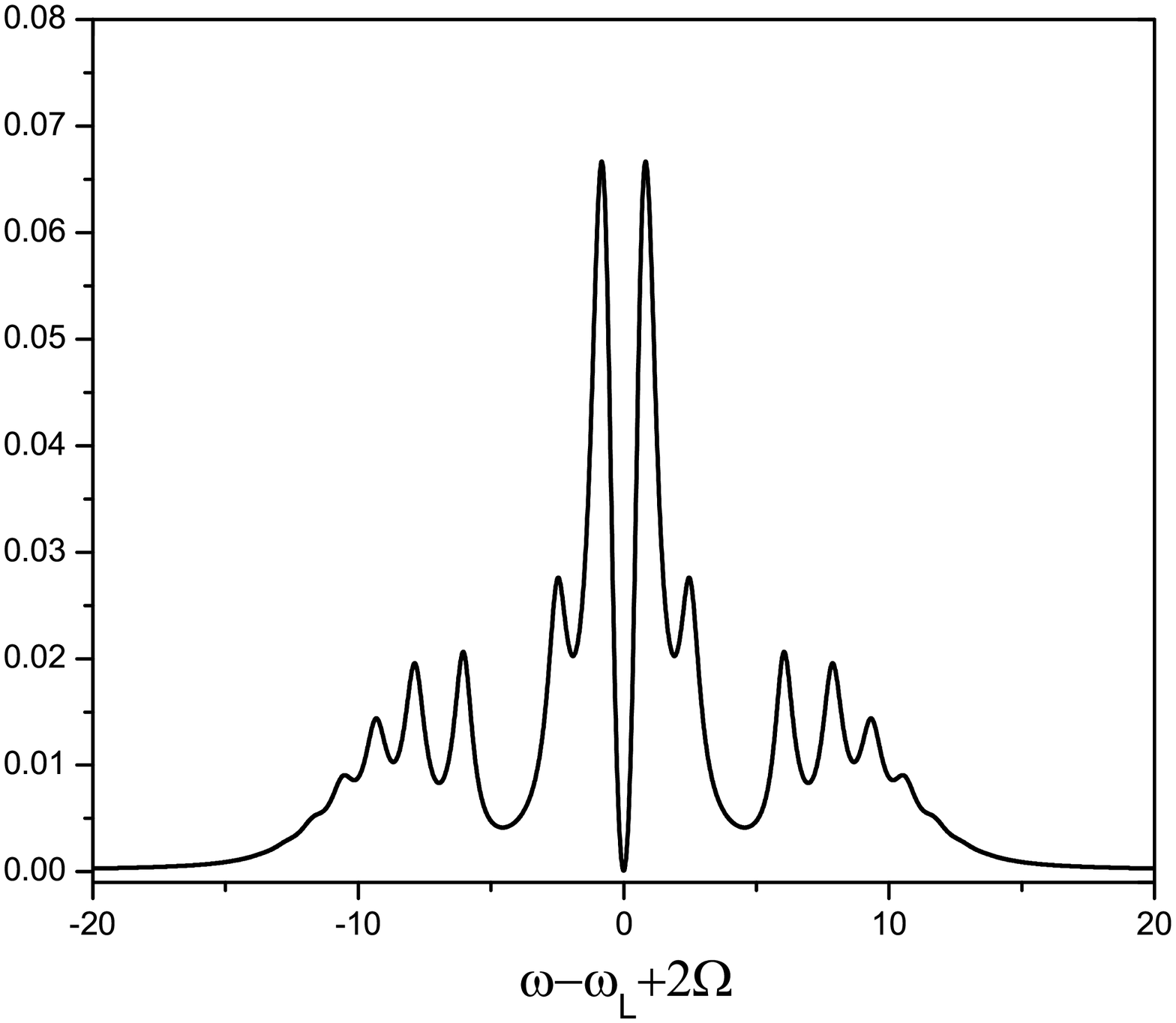}\\
\includegraphics[width=5cm,keepaspectratio,clip]{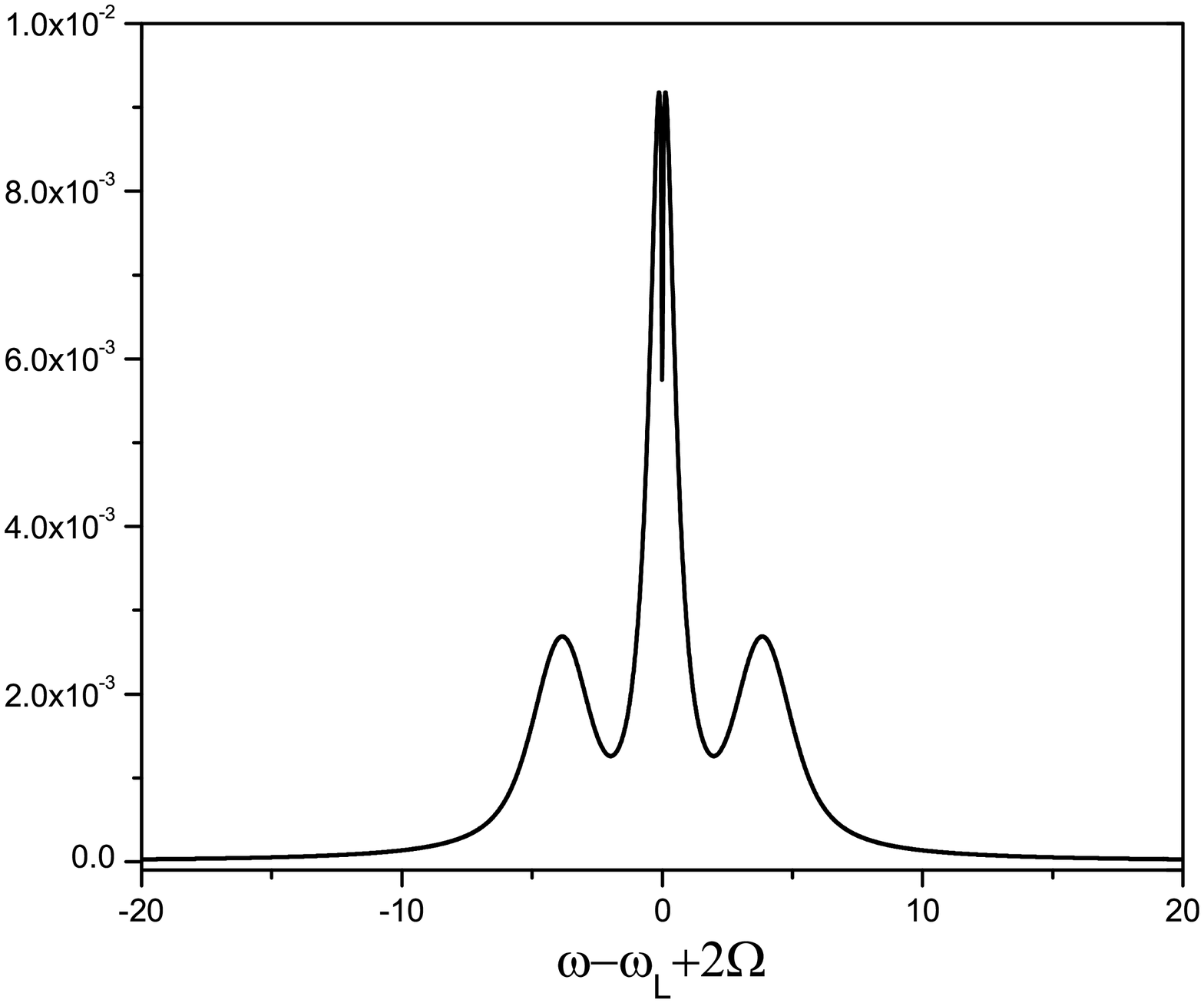}\\
\includegraphics[width=5cm,keepaspectratio,clip]{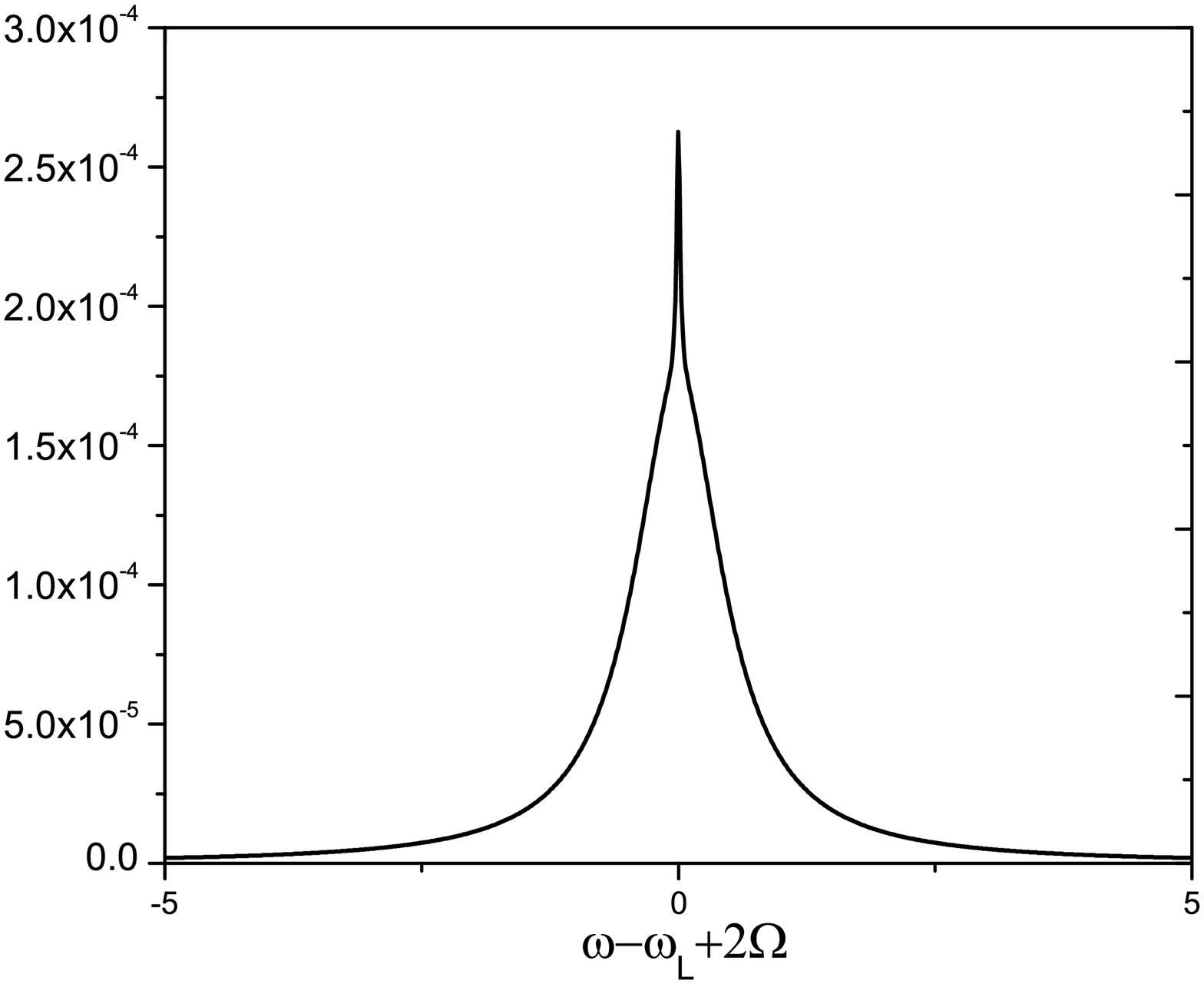}\\
\caption{The atomic fluorescence spectrum $S^{(-)}(\omega)$ as a function of frequency $\omega -\omega_{L}+2\Omega$ for $\kappa = 0.05$, $g=5$, and the band-gap spontaneous emission cancelation $\gamma_{-}= 0$. In frames: (a) $\delta_{a}=-5$, (b) $\delta_{a}= 0$, (c) $\delta_{a}= 1$, (d) $\delta_{a}= 5$. All the parameters are normalized to $\gamma=1$.} \label{fig3}
\end{figure}

\begin{figure}[hbp]
\includegraphics[width=5cm,keepaspectratio,clip]{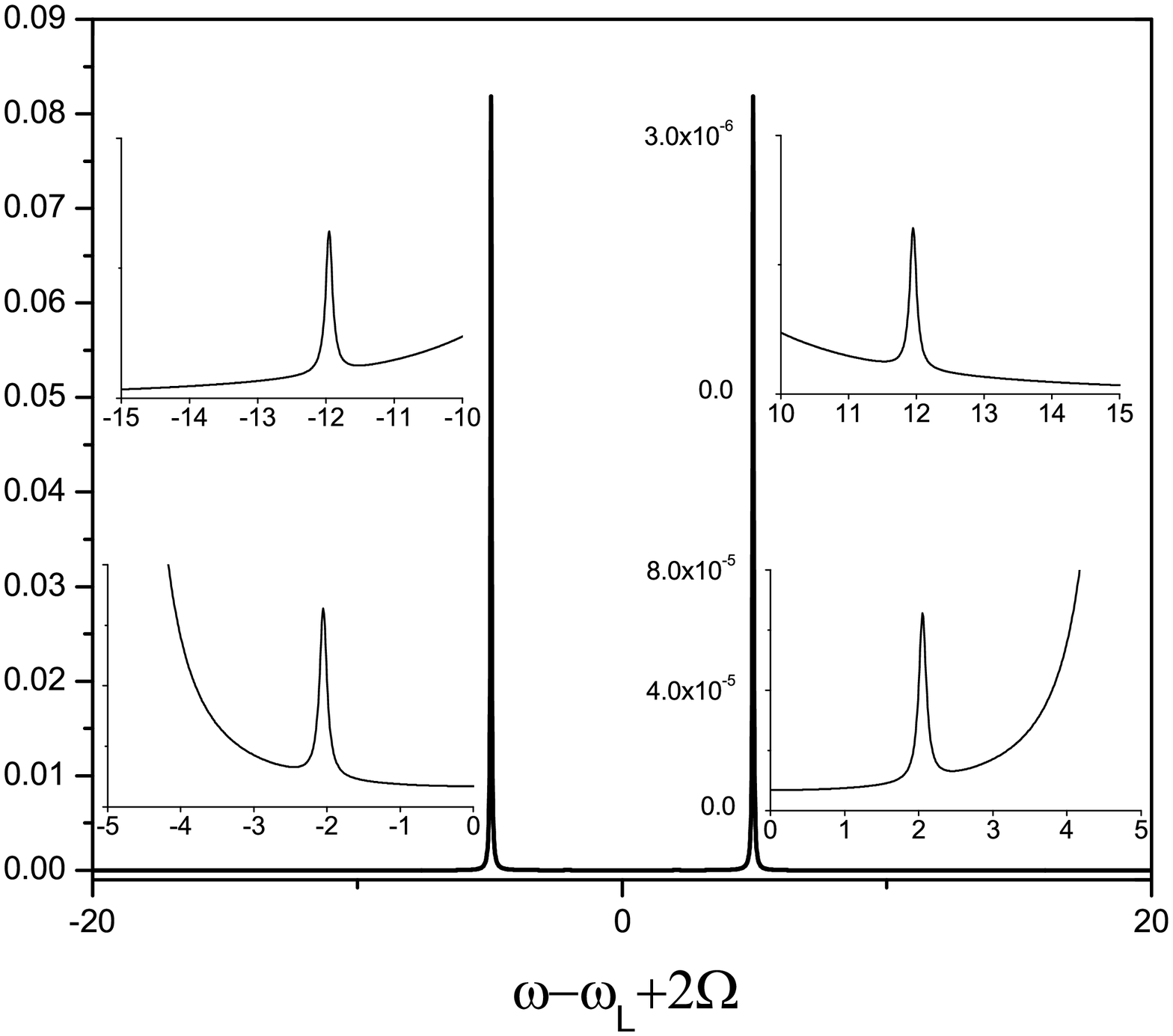}\\
\includegraphics[width=5cm,keepaspectratio,clip]{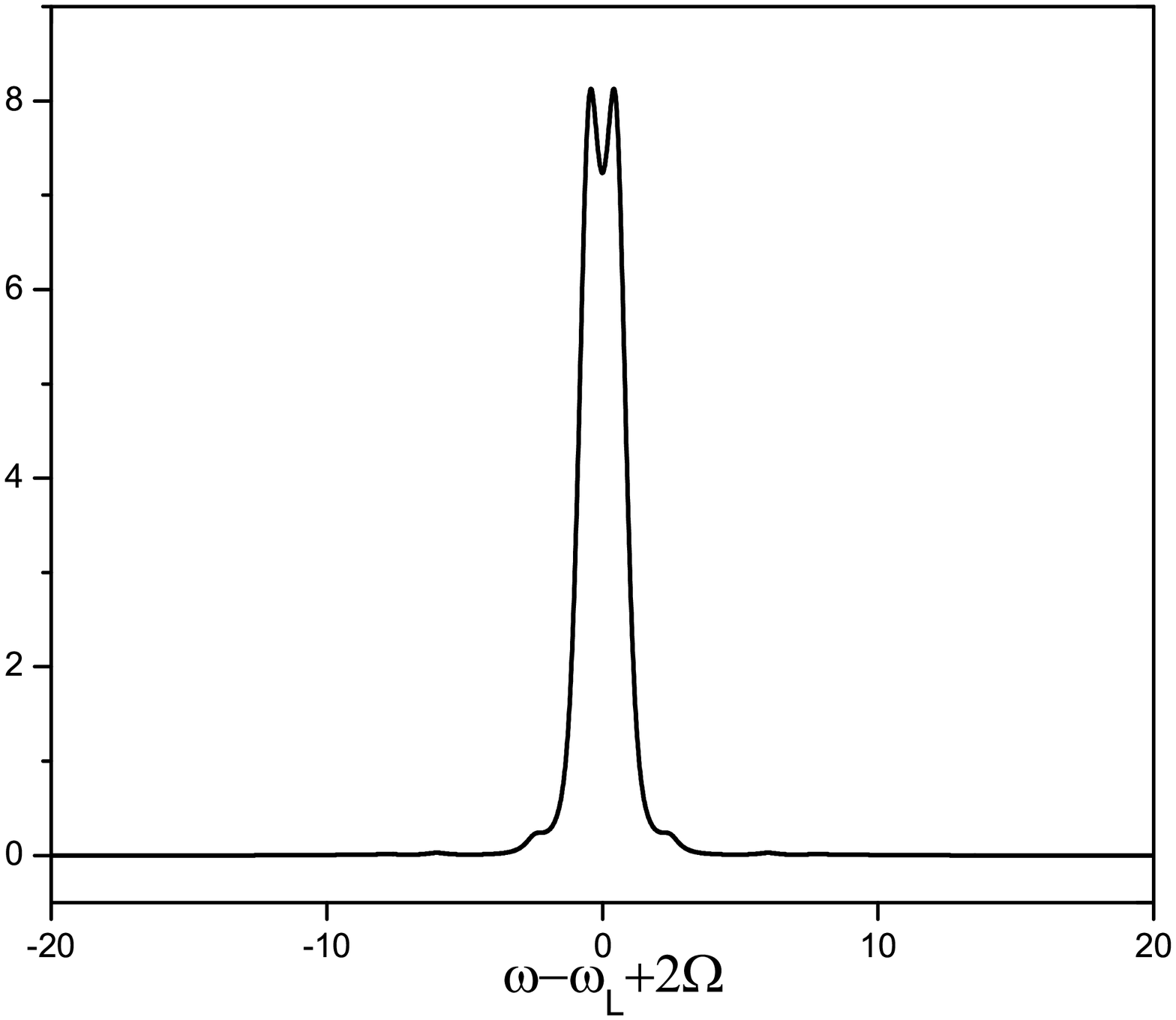}\\
\includegraphics[width=5cm,keepaspectratio,clip]{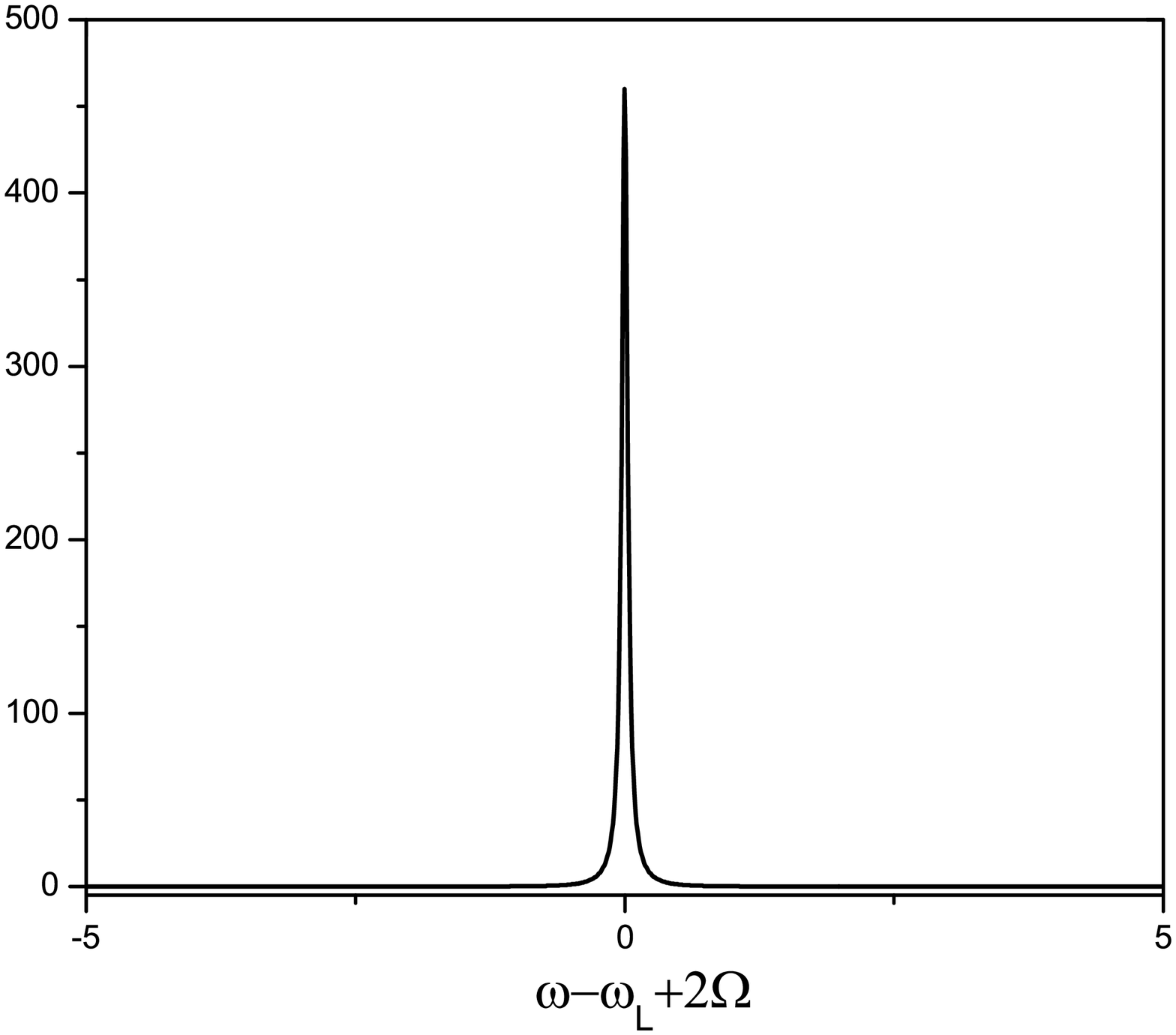}\\
\includegraphics[width=5cm,keepaspectratio,clip]{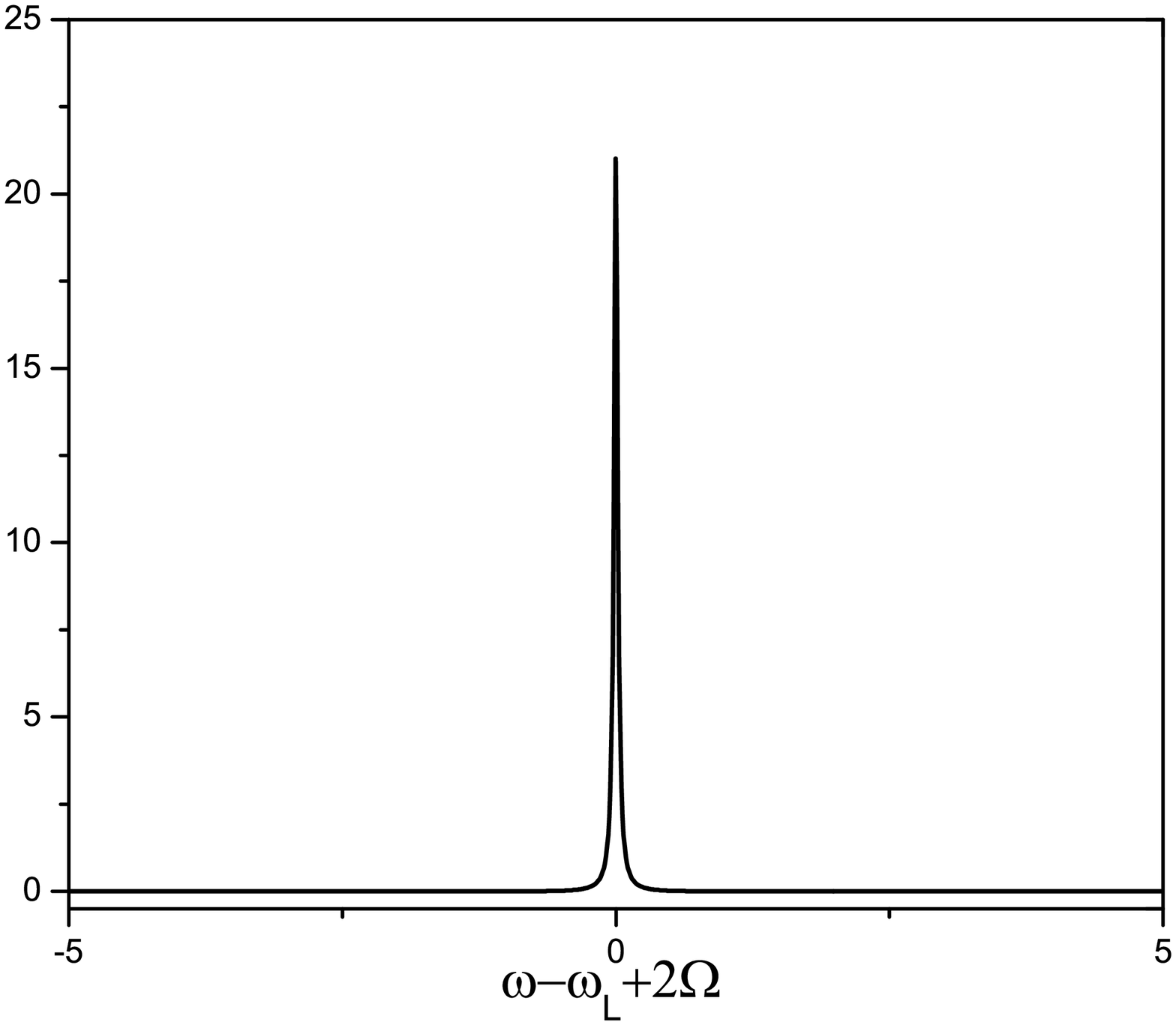}\\
\caption{The cavity field spectrum $S_{c}(\omega)$ as a function of frequency $\omega -\omega_{L}+2\Omega$ for $\kappa = 0.05$, $g=5$, the band-gap spontaneous emission cancelation $\gamma_{-}= 0$. In frames: (a) $\delta_{a}=-5$, (b) $\delta_{a}= 0$, (c) $\delta_{a}= 1$, (d) $\delta_{a}= 5$. All the parameters are normalized to $\gamma=1$.} \label{fig4}
\end{figure}

We now evaluate the atomic fluorescence and the cavity field spectra in a good cavity limit of $g\gg \kappa,\gamma$. We concentrate on the lower Rabi sideband and consider the variation in the spectra with changing $\gamma_{+}$, the pumping rate of the dressed-atom laser for the case of a frequency independent reservoir, $\gamma_{-}\neq 0$ and compare it with the result for the band-gap, $\gamma_{-}=0$ situation. The results are shown in~Figs.~\ref{fig1}-\ref{fig4}.

Let us first examine the spectra in the situation when the atom is coupled to a frequency independent reservoir that the spontaneous transitions are allowed at all three dressed-atom frequencies, i.e. $\gamma_{0}, \gamma_{+}$ and $\gamma_{-}$ are all different from zero. Figure~\ref{fig1} shows the atomic fluorescence spectrum for fixed $\kappa = 0.05\gamma$, and the cavity coupling constant $g=5\gamma$, and for progressively increasing pumping rate $\gamma_{+}$. At low pumping rate, the spectrum  consists of two well-separated narrow peaks, centered at frequencies $\omega = \omega_{-}\pm g_{1}$. This is well known as the vacuum Rabi splitting. As the pumping rate increases, i.e. $\cos^{4}\phi$ increases, more peaks emerge at multiples of~$\pm g_{1}$. At moderate pumping rates, a multi-peaked spectrum is well visible with peaks inside as well as outside the vacuum Rabi doublet, but the most intense are still those corresponding to the vacuum Rabi splitting. Note that there is no fluorescence at $\omega_{-}$, the cavity field frequency. Moreover, the peaks group into three separate sets of multiplets. If we further increase the pumping rate, the multi-peak structures merge into the characteristic Mollow type triplet with the central component at the cavity field frequency $\omega_{-}$ and sidebands located at $\pm 2\sqrt{\langle n\rangle}g_{1}$, where $\langle n\rangle$ is the average number of photons in the cavity mode. This spectrum is very similar to that found before for a bichromatic driving field~\cite{ff96}, with one strong and one weak component. The strong component was resonant with the atomic transition whereas the weaker component was matched  to the Rabi sideband induced by the strong component. In the case considered here, the role of the weaker component is played by the cavity field. 

Somewhat similar effects appear in the cavity field spectrum. Figure~\ref{fig2} shows the cavity field spectrum for the same parameters as in Fig.~\ref{fig1}. We observe that the spectrum is quantitatively similar to the fluorescence spectrum, but there is a qualitative difference that the multi-peak structure builds up {\it only} inside the the vacuum Rabi doublet. When the pumping rate increases further, the multi-peak structure merges into a very marrow peak centered at the cavity frequency $\omega_{-}$. The appearance of the very narrow peak the cavity-field frequency is a clear evidence of a lasing action. Thus, there is a threshold value of $\gamma_{+}$, defined to be that value of the pumping rate for which the cavity field spectrum merges into a very narrow, almost monochromatic peak located at the frequency of the cavity mode.

It is interesting to note from Figs.~\ref{fig1} and~\ref{fig2} that we can distinguish three regions of the pumping rate at which the spectra have completely different structures. The low pumping region at which the spectra exhibit the vacuum Rabi splitting, the moderate region where multi-peak structure is seen, and the region of large pumping rates where only a single very narrow peak is observed.

The behavior of the system and resulting spectral properties can be very different when the atom is coupled to a frequency-dependent reservoir exhibited by a band-gap material.  
Figure~\ref{fig3} illustrates the filtering effect of the band-gap material on the atomic fluorescence spectrum by choosing $\gamma_{-}=0$. It corresponds to the situation of no spontaneous emission on the dressed-atom transition resonant to the cavity-mode frequency. For low pumping rates the behavior of the spectrum is seen to be qualitatively different from the previous case of the frequency independent reservoir. We start to see the multi-peaked structure at very low values of the pumping rate. In addition to a general appearance of the multi-peaked structure at low and moderate pumping rates, an enhancement of the fluorescence and the cavity field appear at frequencies symmetrically displaced from the cavity-mode frequency by the effective coupling constant $g_{1}$. It is notable that in both spectra the resonances occurring at $\pm g_{1}$ are the dominant features in the case of weak and moderate pumping rates. 

Figure~\ref{fig4} shows the behavior of the cavity field spectrum in the band-gap situation with $\gamma_{-}=0$. Again, the multi-peaked structure, which is caused by the population of the low energy entangled states of the combined dressed-atom plus the cavity mode system, is seen to emerge at very low pumping rates. As the pumping rate increases, a threshold behavior occurs that the multi-peak spectrum converts into a single very narrow peak. Simultaneously, the atomic fluorescence spectrum converts into a triplet, similar to the Mollow triplet, with the central component at frequency $\omega_{-}$. It is interesting that the magnitude and the width of the narrow peak is the same as in the case of the frequency independent reservoir, i.e. is independent of whether $\gamma_{-}=0$ or $\gamma_{-}\neq 0$. Thus, the dominant effects for the filtering properties of the band-gap material are see at low and moderate pumping rates.

\section{Interpretation of the results}

The physics underlying the modifications of the spectra in the presence of the band-gap material  can be understood more intuitively by an analytical examination of the energy structure of the combined dressed-atom plus the cavity mode system. The combined states are entangled states between the atom and the cavity field. The knowledge of the states allows us to calculate the populations of these states and transition rates between them. Our treatment is an extension of the Freedhoff and Quang approach to the case of a frequency dependent reservoir. In these calculations, we use the fully quantized dressed-atom approach which gives a complementary view of the cavity effects. 

We use the quantum dressed-atom model, in which we quantize both the driving laser and the cavity-mode fields. In our calculations, we first couple the bare atom to the laser field, then we couple the resulting dressed atom to the cavity mode. In the fully quantum description we replace the semiclassical dressed states by their quantum counterparts 
\begin{eqnarray}
|\tilde{1}, N\rangle = \cos\phi |1,N\rangle- \sin\phi |2,N-1\rangle ,\nonumber\\
|\tilde{2}, N\rangle= \sin\phi |1,N\rangle+ \cos\phi |2,N-1\rangle ,\label{e30}
\end{eqnarray}
where $|i,N\rangle$ is a state in which the atom is in the $i$th state and $N$ photons are present in the driving field. 

The eigenstates of the noninteracting dressed-atom system plus the cavity mode are product states $|\tilde{i},N,n\rangle=|\tilde{i},N\rangle\otimes|n\rangle$, where $\ket n$ is the eigenstate of the cavity mode, with energies
\begin{eqnarray}
E_{i,N,n} = N\omega_{L}+(-1)^{i}\Omega+n(\omega_{L}-2\Omega) ,\label{e31}
\end{eqnarray}
where $n$ is the number of photons in the cavity mode.

It is easily to show that the state $|2,N,0\rangle$ is a single state, whereas the state $|2,N-n,n\rangle$ and $|1,N-n+1,n-1\rangle$ from doubly degenerate pairs.
When we include the interaction between the dressed atom and the cavity mode 
\begin{eqnarray}
V_{ac} = \hbar g\left(a^{\dagger}\sigma_{12}+\sigma_{21}a\right) ,\label{e32}
\end{eqnarray}
the degeneracy is lifted, resulting in doublets with eigenstates, the atom-cavity-mode entangled states of the form
\begin{widetext}
\begin{eqnarray}
\ket{\Psi_{N,\pm n}} = \left\{\begin{array}{ll}
|\tilde{2},N,0\rangle ,&\textrm{for $n={0}$}\\
\\
\frac{1}{\sqrt{2}}\left(|\tilde{2},N-n,n\rangle\pm |\tilde{1},N-n+1,n-1\rangle\right) , &\textrm{for $n\neq{0}$} 
\end{array}\right. \label{e33}
\end{eqnarray}
\end{widetext}
corresponding to energies
\begin{align}
E_{N,0} &= \hbar\left(N\omega_{L}+\Omega\right) ,\nonumber\\
E_{N,\pm{n}} &= \hbar\left[N\omega_{L} -(2n-1)\Omega\pm{g_1}\sqrt{n}\,\right] .\label{e34}
\end{align}
The entangled states (\ref{e33}) are the energy states of the system. They group into manifolds, each containing a singlet and an infinite number of doublets. Neighboring doublets are separated by~$2\Omega$, while the intra-doublet states are separated by $2\sqrt{n}g_{1}$. The spectral properties of the atomic fluorescence and the cavity field are a signature of the entangled system, its energy levels, their populations and probabilities of spontaneous transitions between them. Thus, one would expect multiple structures in the spectra.

\subsection{Transition rates}

The energy states of the system interact with the vacuum reservoir modes. This results in a spontaneous emission cascade by the system down its energy manifold ladder. The probability of a transition between any two energy states is proportional to the absolute square of the dipole transition moment between them. Since the interaction between the atom and the field is determined by a single-photon Hamiltonian, the transition dipole moments have selection rules that nonzero transitions can occur only between states $\ket{\Psi_{N,\pm n}}$ and $\ket{\Psi_{N-1,\pm n^{\prime}}}$ with $n^{\prime} = n, n\pm 1$. The probabilities (spontaneous emission rates), corresponding to transitions at frequencies
\begin{eqnarray}
\omega_{\pm n,n^{\prime}} = \hbar^{-1}\left(E_{N,\pm n} - E_{N-1,n^{\prime}}\right) ,\label{e35}
\end{eqnarray}
are given by
\begin{align}
\Upsilon_{\pm{n},\pm{n}^{\prime}} &= \frac{1}{4}\gamma_{0}|\bra{\Psi_{N,\pm n}}R_{3}\ket{\Psi_{N-1,\pm n^{\prime}}}|^{2} \nonumber\\
&+\gamma_{+}|\bra{\Psi_{N,\pm n}}R_{21}\ket{\Psi_{N-1,\pm n^{\prime}}}|^{2} \nonumber\\
&+\gamma_{-}|\bra{\Psi_{N,\pm n}}R_{12}\ket{\Psi_{N-1,\pm n^{\prime}}}|^{2} .\label{e36}
\end{align}
Similarly, cavity damping cause transition to occur from states
$|N,\pm{n}\rangle$ to states $|N-1,\pm{n}^{\prime}\rangle$ with probabilities
\begin{eqnarray}
\kappa_{\pm{n},\pm{n}^{\prime}} = \kappa|\bra{\Psi_{N,\pm n}}a^{\dagger}\ket{\Psi_{N-1,\pm n^{\prime}}}|^{2} .\label{e37}
\end{eqnarray}

Since our interest is principally on the spectral distribution at the frequency of the lower Rabi sideband, $\omega_{-}=\omega_{L}-2\Omega$, we therefore focus on spontaneous emission rates between states $\ket{\Psi_{N,\pm{n}}}\rightarrow\ket{\Psi_{N-1,\pm{(n+1)}}}$ and $\ket{\Psi_{N,\pm{(n+1)}}}\rightarrow\ket{\Psi_{N-1,\pm{n}}}$.
Using Eqs.~(\ref{e33}), (\ref{e34}) and (\ref{e36}), we readily find that the spontaneous emission rates corresponding to several frequencies centered at $\omega_{-}$ are given by the expressions
\begin{align}
\Upsilon_{\pm{n},\pm{(n+1)}} &= \frac{1}{4}\gamma_{+} (1+\delta_{n,0}) ,\ \omega_{\pm{n},\pm{(n+1)}} = \omega_{-}\mp \nu_{n}^{(-)} ,\nonumber\\
\Upsilon_{\mp{n},\pm{(n+1)}} &=  \frac{1}{4} \gamma_{+} (1+\delta_{n,0}) ,\ \omega_{\mp{n},\pm{(n+1)}} = \omega_{-}\mp \nu_{n}^{(+)} ,\nonumber\\
 \Upsilon_{\pm{(n+1)},\pm{n}} &=\frac{1}{4}\gamma_{-}(1+\delta_{n,0}) ,\ \omega_{\pm
(n+1),\pm{n}} = \omega_{-}\pm \nu_{n}^{(-)} ,\nonumber\\
\Upsilon_{\pm{(n+1)},\mp{n}} &= \frac{1}{4}\gamma_{-}(1+\delta_{n,0}) ,\ \omega_{\pm (n+1),\mp{n}} = \omega_{-}\pm \nu_{n}^{(+)} ,\label{e38}
\end{align}
where $\nu_{n}^{(\pm)} = \left(\sqrt{n+1}\pm\sqrt{n}\,\right)\!g_{1}$.

Similarly, we find from Eqs.~(\ref{e33}), (\ref{e34}) and (\ref{e37}) that the cavity damping causes transitions with the rates
\begin{align}
\kappa_{\pm{(n+1)},\pm{n}} &=\frac{1}{4}\kappa \left(\sqrt{n+1}+\sqrt{n}\, \right)^2 , \ \omega_{\pm (n+1),\pm{n}} = \omega_{-}\pm \nu_{n}^{(-)}  \nonumber\\
\kappa_{\pm{(n+1)},\mp{n}} &=\frac{1}{4}\kappa \left(\sqrt{n+1}-\sqrt{n}\, \right)^2 ,\ \omega_{\pm (n+1),\mp{n}} = \omega_{-}\pm \nu_{n}^{(+)} .\label{e38a}
\end{align}

From the transition rates and the corresponding transition frequencies, it is apparent that the spectra consist of multiple lines located symmetrically about $\omega_{-}$. The multiples split into two sets of lines, $\nu_{n}^{(-)}$ and $\nu_{n}^{(+)}$, which we call the inner and outer sidebands, respectively. The spontaneous emission rates are dependent on the density of the vacuum reservoir modes at both the higher and lower Rabi sideband. However, only the rates that depend on the density of the modes at the lower Rabi sideband are, in addition, affected by the cavity damping. Moreover, the effect of the cavity damping on the transition rates corresponding to the inner sidebands increase with increasing number of photons $n$, whereas the transition rates corresponding to the outer sideband decreases with increasing $n$. This results in the enhancement (reduction) the inner (outer) sidebands amplitudes with the increasing pumping rate, as seen in Figs.~\ref{fig2} and \ref{fig4}. 
It  is interesting to note that there are no spontaneous transitions allowed at the cavity field frequency $\omega_{-}$ unless $n\gg 1$ for which $\sqrt{n+1}-\sqrt{n}\approx 0$. Consequently, there is no spectral line at $\omega_{-}$, the cavity mode frequency, until a large number of photons accumulate in the cavity mode, $n\gg 1$. Then, the inner sidebands at $\pm\nu_{n}^{(-)}$ merge into a single peak at $\omega_{-}$, while the outer sidebands merge into single lines located at frequencies $\pm 2\sqrt{n}g_{1}$. In this limit, the atomic fluorescence spectrum exhibits three lines. The merging of the fluorescence spectrum into the familiar Mollow triplet is a clear evidence that in this limit the cavity field appears as a monochromatic coherent field. We could call the point at which the cavity field merges into a single narrow peak as a threshold for the cavity field to operate as a conventional laser. On the other hand, the cavity field spectrum converts into a single peak, that the outer sidebands decreases with increasing $n$ with the inner sidebands exhibiting a strong enhancement. This is consistent with the variation with $n$ of the cavity induced damping rates~(\ref{e38a}).

\subsection{Populations}

To evaluate the populations of the energy states of the system, we project the master 
equation~(\ref{e14}) onto $\ket{\Psi_{N,\pm n}}$ on the right and $\bra{\Psi_{N,\pm n}}$ on the left, and make the secular approximation that ignores couplings between populations and coherences. We then introduce the reduced populations
\begin{eqnarray}
\Pi_{\pm n} = \sum_{N} \bra{\Psi_{N,\pm n}} \rho \ket{\Psi_{N,\pm n}} ,\label{e39}
\end{eqnarray}
and find that in the steady state, $\Pi_{\pm n}$ satisfy the recurrence relation
\begin{eqnarray}
&&\gamma_{+}{\Pi}_{n-1} - \left[\gamma_{+}+\gamma_{-}+(2n-1)\kappa\right]\Pi_{n}\nonumber\\
&&+\left[\gamma_{-}+(2n+1)\kappa\right]\Pi_{n+1} = 0 ,\label{e40}
\end{eqnarray}
with $(\gamma_{-}+\kappa)\Pi_{1} = \gamma_{+}{\Pi}_{0}$ and ${\Pi}_{-n}={\Pi}_{n}$.

The solution of Eq.~(\ref{e36}) gives
\begin{align}
{\Pi}_{\pm{n}}=\Pi_{0}\prod^{n}_{m=1}\frac{\gamma_{+}}{\gamma_{-}+(2m-1)\kappa} ,\label{e41}
\end{align}
where $\Pi_{0}$ is determined by the normalization condition.

It is evident from Eq.~(\ref{e41}) that the population of the energy states depends solely on the density of the vacuum modes at the Rabi sideband frequencies. The population is unequally distributed among the states, but the redistribution depends strongly on whether $\gamma_{-}=0$ or~$\gamma_{-}\neq 0$. For $\gamma_{-}\neq 0$ and $\kappa\ll\gamma$, the population distribution depends weakly on $n$ leaving the state $n>1$ practically unpopulated. This explains why in the limit of a low pumping, $\gamma_{+}\ll 1$, only the characteristic vacuum Rabi splitting is seen in the spectra, as shown in Figs.~\ref{fig1} and~\ref{fig2}. However, in the band-gap situation of $\gamma_{-}=0$ the population distribution is strongly sensitive to $n$ and the population can be redistributed over the $n>1$ state even for relatively weak pumping rate. This clearly explains why in the band-gap situation, shown in Figs.~\ref{fig3} and~\ref{fig4}, a multi-peak structure is observed in the spectra even for very low pumping rate.

\section{Conclusions}

We have examined the atomic fluorescence and cavity field spectra of a strongly driven two-level atom coupled to a single-mode cavity and a frequency-dependent reservoir exhibited by a band-gap material. The band-gap spectral density of the electromagnetic vacuum modes changes sharply at the band-edge frequency $\omega_{b}$, such that $\omega_{b}-\omega_{-} \gg \Gamma$, where $\omega_{-}$ is the frequency of the lower Rabi sideband and $\Gamma$ is its bandwidth. 
We have found that the frequency dependent reservoir affects the spectra for low and moderate pumping rates, while for large pumping rates the spectra do not differ substantially from that predicted for the case of a frequency independent reservoir.
The effect of the frequency dependent reservoir is to narrow the spectral lines and to suppress the below threshold regime characterized by the vacuum Rabi splitting. Multi-peaked spectra are observed for essentially very low pumping rates. This feature can be regarded as a signature of the thresholdless lasing. However, the multi-peaked structure of the cavity field spectrum indicates that the system does not operate as a source of a monochromatic light.  
A conventional laser is usually described as a source of a highly monochromatic coherent light.
We have found that the cavity field spectrum exhibits a threshold behavior that at a certain pumping rate, the multi-peak spectrum converts into a single very narrow peak located at the cavity mode frequency. The threshold behavior exists despite the fact that there is no threshold behavior in the photon statistics of the cavity field.

\section*{Acknowledgments} 

We acknowledge financial support from the National Natural Science Foundation of China (Grant Nos. 10674052 and 60878004), the Ministry of Education under project NCET (grant no NCET-06-0671), SRFDP (under grant no. 200805110002), and the National Basic Research Project of China (grant no. 2005 CB724508).

\appendix

\section{}

In this Appendix, we give the explicit expressions for the vector $Z^{(m)}_{n}$ and for the matrices of the coefficients appearing in Eq.~(\ref{e18}):
\begin{align}
{\bf Z}^{(m)}_{n} &=\left(\begin{array}{ccc}
\rho^{(1)}_{n,n+m}\\\rho^{(2)}_{n,n+m}\\\rho^{(3)}_{n,n+m}\\\rho^{(4)}_{n,n+m}\end{array}\right) ,
\end{align}

\begin{align}
{\bf A}^{(m)}_{n} &= \frac{1}{2}\alpha_{n,m}\left(\begin{array}{cccc}
0&0&0&0\\
0&0&0&0\\
-g_{1}&g_{1}&0&0\\
0&0&0&0
\end{array}\right) ,
\end{align}

\begin{widetext}
\begin{align}
{\bf B}^{(m)}_{n} &= \left(\begin{array}{cccc}
-\kappa\beta_{n,m}&0&-2g_1&2g_1\\
-(\gamma_{+}-\gamma_{-})&-[(\gamma_{+}+\gamma_{-})+\kappa\beta_{n,m}]&-2g_1&-2g_1\\
\frac{1}{2}g_{1}\beta_{n,m}&\frac{1}{2}g_{1}\beta_{n,m}&-\Gamma_{c} -\kappa\left(\beta_{n,m}-\frac{1}{2}\right)&-\kappa\\
-\frac{1}{2}g_{1}\beta_{n+1,m}&\frac{1}{2}g_{1}\beta_{n+1,m}&0&-\Gamma_{c} -\kappa\left(\beta_{n,m}+\frac{1}{2}\right)
\end{array}\right) ,
\end{align}
\end{widetext}

\begin{align}
{\bf C}^{(m)}_{n} &= \alpha_{n+1,m}\left(\begin{array}{cccc}
\kappa &0&0&0\\
0&\kappa &0&0\\
0&0&\kappa &0\\
\frac{1}{2}g_{1} &\frac{1}{2}g_{1}&0&\kappa
\end{array}\right) ,
\end{align}
where $\alpha_{n,m} =\sqrt{n(n+m)}$ and $\beta_{n,m} =n+m/2$.


\begin{thebibliography}{99}

\bibitem{sz97} Scully, M. O.; Zubairy, M. S. {\it Quantum Optics}, (Cambridge University Press, Cambridge, 1997).
\bibitem{m69} Mollow, B. R. {\em Phys. Rev.} {\bf 1969}, {\em 188}, 1969. 
\bibitem{km76} Kimble, H. J.; Mandel, L. {\em Phys. Rev.} {\bf 1976}, {\em A13}, 2123.
\bibitem{be94} For a review see {\em Cavity Quantum Electrodynamics}, edited by Berman, (Academic Press, Boston, 1994).
\bibitem{ct92} Cohen-Tannoudji, C.; Dupont-Roc, J.; Grynberg, G. {\bf 1992}, {\em Atom-Photon
Interactions}, (Wiley, New York).
\bibitem{sm83} Sanchez-Mondragon, J. J.; Narozhny, N. B.; Eberly, J. H. {\em Phys. Rev. Lett.} {\bf 1983}, {\em 51}, 550.
\bibitem{ag84} Agarwal, G. S. {em Phys. Rev. Lett.} {\bf 1984}, {\em 53}, 1732.
\bibitem{cb89} Carmichael, H. J.; Brecha, R. J.; Raizen, M. G.; Kimble, H. J.; Rice, P. R. 
{\em Phys. Rev.} {\bf 1989}, {\em A40}, 5516.
\bibitem{rt91} Rempe, G.; Thompson, R. J.; Brecha, R. J.; Lee, W. D.; Kimble, H. J. {\em Phys. Rev. Lett.} {\bf 1991}, {\em 67}, 1727.
\bibitem{th95} Turchette, Q. A.; Hood, C. J.; Lange, W.; Mabuchi, H.; Kimble, H. J. {\em Phys. Rev. Lett.} {\bf 1995}, {\em 75}, 4710.
\bibitem{gw92} Gauthier, D. J.; Wu, Q.; Morin, S. E.; Mossberg, T. W. {\em Phys. Rev. Lett.} {\bf 1992}, {\em 68}, 464.
\bibitem{qf93} Quang, T.; Freedhoff, H. {\em Phys. Rev.} {\bf 1993}, {\em A47}, 2285.
\bibitem{lg97} L\"offler, M.; Meyer, G. M.; Walther, H. {\em Phys. Rev.} {\bf 1997}, {\em A55}, 3923.
\bibitem{fq93} Freedhoff, H.; Quang, T. {\em J. Opt. Soc. Am.} {\bf 1993}, {\em B10}, 1337. 
\bibitem{nso90} Narducci, L. M.; Scully, M. O.; Oppo, G. L.; Ru, P.; Tredicce, J. R. {\em Phys. Rev.} {\bf 1990}, {\em A42}, 1630.
\bibitem{gzm91} Gauthier, D. J.; Zhu, Y.; Mossberg, T. W. {\em Phys. Rev. Lett.} {\bf 1991}, {\em 66}, 2460.
\bibitem{kk95} Keitel, C. H.; Knight, P. L.; Narducci, L. M.; Scully, M. O. {\em Optics Commun.} {\bf 1995}, {\em 118}, 143.
\bibitem{lm88} Lewenstein, M.; Mossberg, T. W. {\em Phys. Rev.} {\bf 1988}, {\em A37}, 2048. 
\bibitem{zl91} Zakrzewski, J.; Lewenstein, M.; Mossberg, T. W. {\em Phys. Rev.} {\bf 1991}, {\em  A44}, 7717, 7732, 7746.
\bibitem{lz90} Lewenstein, M.; Zhu, Y.; Mossberg, T. W. {\em Phys. Rev. Lett.} {\bf 1990},  {\em 64}, 3131.
\bibitem{lb91} Lu, N.; Berman, P. R. {\em Phys. Rev.} {\bf 1991}, {\em A44}, 5965. 
\bibitem{fj04} Florescu, L.; John, S.; Quang, T.; Wang, R. {\em Phys. Rev.} {\bf 2004}, {\em A69}, 013816.
\bibitem{fl06} Florescu, L. {\em Phys. Rev.} {\bf 2006}, {\em A74}, 063828. 
\bibitem{tl08} Tan, R,; Li, G.-X.; Ficek, Z. {\em Phys. Rev.} {\bf 2008}, {\em A78}, 023833.
\bibitem{ll09} Li, G.-X.; Luo, M.; Ficek, Z. {\em Phys. Rev.} {\bf 2009}, {\em A79}, 053847. 
\bibitem{rc94} Rice, P. R.; Carmichael, H. J. {\em Phys. Rev.} {\bf 1994}, {\em A50}, 4318.
\bibitem{tp94} Pellizzari, T.; Ritsch, H. {\em Phys. Rev. Lett.} {\bf 1994}, {\em 72}, 3973.
\bibitem{dm88} De Martini, F.; Jacobovitz, G. R. {\em Phys. Rev. Lett.} {\bf 1988}, {\em 60}, 1711.
\bibitem{bk94} Bj\"ork, G.; Karlsson, A.; Yamamoto, Y. {\em Phys. Rev.} {\bf 1994}, {\em A50}, 1675.
\bibitem{pd99} Protsenko, I.; Domokos, P.; Lefevre-Seguin, V.; Hare, J.; Raimond, J. M.;  Davidovich, L. {\em Phys. Rev.} {\bf 1999}, {\em A59}, 1667.
\bibitem{sh99} Szymanska, M. H.; Hughs, A. F.; Pike, E. R. {\em Phys. Rev. Lett.} {\bf 1999}, {\em 83}, 69.
\bibitem{wj04} Wang, R.; John, S. {\em Phys. Rev.} {\bf 2004}, {\em A70}, 043805.
\bibitem{ya87} Yablonovitch, E. {\em Phys. Rev. Lett.} {\bf 1987}, {\em 58}, 2059.
\bibitem{jo87} John, S. {\em Phys. Rev. Lett.} {\bf 1987},  {\em 58}, 2486.
\bibitem{jq94} John, S.; Quang, T. {\em Phys. Rev.} {\bf 1994}, {\em  A50}, 1764.
\bibitem{ff96} Ficek, Z.; Freedhoff, H. S. {\em Phys. Rev.} {\bf 1996}, {\em 53}, 4275. 

\end{thebibliography}
\end{document}